\documentclass[twocolumn,showpacs,preprintnumbers,amsmath,amssymb]{revtex4-1}
\usepackage{graphicx} 
\usepackage{dcolumn} 
\usepackage{bm} 
\usepackage{braket} 
\usepackage{multirow} 
\usepackage[usenames,dvipsnames]{color} 
\usepackage[
colorlinks=true, citecolor=blue, urlcolor=blue, linkcolor=blue,
setpagesize=false, bookmarks=false
]{hyperref}

\usepackage[english]{babel}
\addto{\captionsenglish}{}

\usepackage[resetlabels]{multibib}
\newcites{SM}{Refs in Supplemental Material}

\begin{document}


\title{
Bulk Photovoltaic Effect Driven by Collective Excitations in a Correlated Insulator
}
\author{
Tatsuya Kaneko$^{1}$, Zhiyuan Sun$^1$, Yuta Murakami$^2$, Denis Gole\v{z}$^{3,4,5}$, and Andrew J. Millis$^{1,3}$
}
\affiliation{
$^1$Department of Physics, Columbia University, New York, New York 10027, USA \\
$^2$Department of Physics, Tokyo Institute of Technology, Meguro, Tokyo 152-8551, Japan\\
$^3$Center for Computational Quantum Physics, Flatiron Institute, New York, New York 10010, USA\\
$^4$Faculty of Mathematics and Physics, University of Ljubljana, Jadranska 19, SI-1000 Ljubljana, Slovenia\\
$^5$Jozef Stefan Institute, Jamova 39, SI-1000 Ljubljana, Slovenia
}
\date{\today}


\begin{abstract}
We investigate the bulk photovoltaic effect, which rectifies light into electric current, in a collective quantum state with correlation driven electronic ferroelectricity. 
We show via explicit real-time dynamical calculations that the effect of the applied electric field on the electronic order parameter leads to a strong enhancement of the bulk photovoltaic effect relative to the values obtained in  a conventional insulator.   
The enhancements include both resonant enhancements at sub-band-gap frequencies, arising from excitation of optically active collective modes, and broadband enhancements arising from nonresonant deformations of the electronic order. 
The deformable electronic order parameter produces an injection current contribution to the bulk photovoltaic effect that is entirely absent in a rigid-band approximation to a time-reversal symmetric material. 
Our findings establish that correlation effects can lead to the bulk photovoltaic effect and demonstrate that the collective behavior of ordered states can yield large nonlinear optical responses. 
\end{abstract}

\maketitle


The photovoltaic effect is the optical process that converts light into electrical current~\cite{belinicher1980,vonbalz1981,sipe2000}. 
Photovoltaic effects can be obtained from devices with interfaces (e.g., $p$-$n$ junctions), but potential applications to new types of solar cells have driven  recent  interest  in  bulk photovoltaic effects (BPVE) occurring in homogeneous noncentrosymmetric materials, where artificially fabricated interfaces are not required and a photovoltage is not limited by a band gap energy~\cite{tan2016,spanier2016,cook2017,rangel2017,zhang2019}.

The BPVE has been extensively analyzed in ferroelectrics \cite{koch1975,young2012,choi2009,young2012BFO,nakamura2017,ogawa2017,sotome2019} and Weyl semimetals~\cite{zhang2018,ma2019,osterhoudt2019,ahn2020}. 
Theory distinguishes shift and injection current contributions to the BPVE, generated by photoinduced changes in the electron position and velocity, respectively~\cite{sipe2000,ahn2020}.  
The shift current contribution in ferroelectrics and Weyl semimetals has been related to Berry connection and  topological effects~\cite{morimoto2016,fregoso2017,nagaosa2017,ahn2020}. 
The fundamental assumption underlying this previous work is that the BPVE may be studied in a model of independent electrons moving in a rigid band structure. 
This assumption has important consequences. 
For example, while an injection current contribution has the potential to lead to large nonlinear conductivity~\cite{zhang2019}, the independent particle approximation predicts that the injection current under linearly polarized light vanishes in time-reversal-symmetric insulators~\cite{sipe2000,ahn2020}. 

In this Letter, we show that electronic correlation effects can substantially enhance the BPVE, opening a new pathway to the design of optoelectronic materials. 
The crucial new point is that if the inversion symmetry breaking arises from a low energy electronic instability, then in addition to resonant enhancements at electronic collective mode frequencies, an applied electric field can nonresonantly deform the electronic band structure in ways that  activate an injection current contribution.

We investigate the effect theoretically in the context of the  excitonic insulator (EI) but we emphasize that our generic results apply to any inversion-symmetry breaking collective electronic states. 
Recently proposed EI candidate materials include TiSe$_2$~\cite{cercellier2007,monney2009,kogar2017,kaneko2018}, Ta$_2$NiSe$_5$~\cite{wakisaka2009,kaneko2013,*kaneko2013e,seki2014,lu2017,sugimoto2018,matsubayashi2021}, and WTe$_2$~\cite{wang2021,lee2021,jia2020}. 
The EI state is characterized by  spontaneous band hybridization triggered by the  interband Coulomb interaction in narrow-gap semiconductors and semimetals~\cite{keldysh1965,jerome1967,halperin1968,littlewood2004,bronold2006,kunes2015}. 
The excitonic order may break inversion symmetry, leading to ``electronic ferroelectricity''~\cite{batyev1980,portengen1996,batista2002,sun2021}, in which case the state is referred to as a ferroelectric EI (FEI). 
In this Letter, we investigate the BPVE in a FEI.     
We compute the optical response including changes in the order parameter, identify the optically active collective modes in the FEI, and demonstrate that nonlinear excitation leads  not only to resonant enhancement of the shift  current  but also  to a nonvanishing injection current with both resonant and broadband contributions. 
The nonvanishing injection current contribution means that in contrast to simple band insulators the FEI can exhibit large nonlinear conductivity.

\begin{figure}[t]
\begin{center}
 \begin{tabular}{cc}
 \begin{minipage}[t]{0.45\hsize}
 \centering
 \includegraphics[width=\columnwidth]{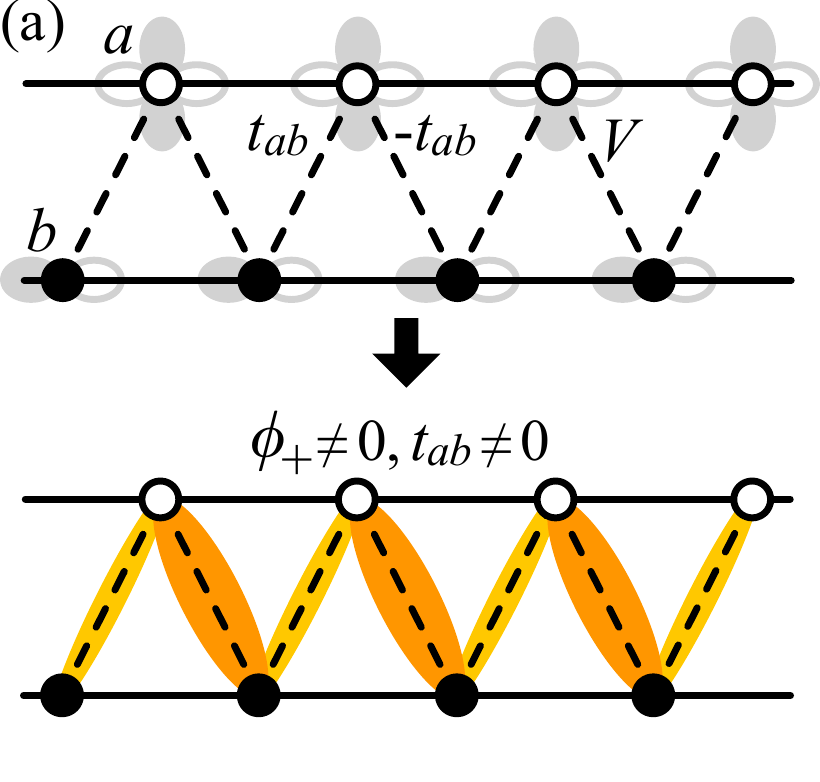}
 \end{minipage} &
 \begin{minipage}[t]{0.45\hsize}
 \centering
 \includegraphics[width=\columnwidth]{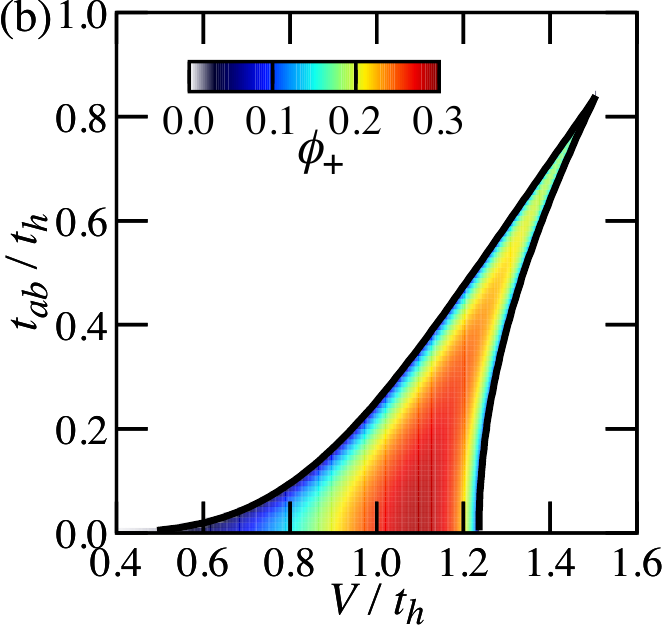}
 \end{minipage}
\end{tabular}
\caption{(a) Top panel: Zigzag chain model Eq.~(\ref{H}), where we plot an example of two orbitals that lead to hopping $t_{ab}$. 
Bottom panel: schematic picture of the EI state. 
(b) Ground-state phase diagram of  Eq.~(\ref{H}) in the plane of  interaction $V$ and interchain hopping $t_{ab}$ computed in the Hartree-Fock approximation for $D/t_h=1$. 
The magnitude of the order parameter $\phi_+$ is shown in false color. 
}
\label{fig1}
\end{center}
\end{figure}

A minimal theoretical model of the FEI [Fig.~\ref{fig1}(a)] is  
\begin{align}
\hat{\mathcal{H}} \!=\! 
&- t_a \! \sum_{j}  \! \left( \hat{c}^{\dag}_{j+1,a} \hat{c}_{j,a} \!+\! {\rm H.c.} \right)
- t_b \! \sum_{j} \! \left( \hat{c}^{\dag}_{j+1,b} \hat{c}_{j,b} \!+\! {\rm H.c.} \right)
 \notag \\
&- t_{ab} \! \sum_{j} \!  \left( \hat{c}^{\dag}_{j,a} \hat{c}_{j,b} \!+\! {\rm H.c.} \right) 
+ t_{ab} \! \sum_{j} \! \left( \hat{c}^{\dag}_{j,a} \hat{c}_{j-1,b} \!+\! {\rm H.c.}  \right)
\notag \\
&+ D \sum_{j}  \left( \hat{n}_{j,a} \!-\! \hat{n}_{j,b} \right)
+V\sum_{j}  \hat{n}_{j,a} \left( \hat{n}_{j,b} \!+\! \hat{n}_{j-1,b} \right)  ,
\label{H}
\end{align} 
where $ \hat{c}_{j,\alpha}$ ($ \hat{c}^{\dag}_{j,\alpha}$) is the annihilation (creation) operator of a fermion on the chain (orbital) $\alpha$ $(=a, b)$ at site $j$, and $\hat{n}_{j,\alpha} = \hat{c}^{\dag}_{j,\alpha} \hat{c}_{j,\alpha} $.  
$t_{\alpha}$ is the hopping integral on the chain ${\alpha}$, and $t_{ab}$ is the interchain hopping that has the opposite signs along the $+x$ and $-x$ directions.    
This type of hopping appears when  two orbitals have opposite parities along the chain direction [see, e.g., Fig.~\ref{fig1}(a)]~\cite{sandu2005,mazza2020}.   
$D$ is the energy level difference, and $V$ is the interchain Coulomb interaction that induces the excitonic instability.    
For simplicity, we take a  particle-hole symmetric band structure with $t_{a} = - t_{b} = t_h$ but our results do not depend in any important way on this assumption.  
We  focus on the half-filled case $\braket{\hat{n}_{j,a}} + \braket{\hat{n}_{j,b}}=1$ and set $t_h$ ($t_h^{-1}$) as a unit of energy (time)~\cite{ceha}.  

Excitonic order in Eq.~(\ref{H}) is characterized by the expectation values  $\phi(x/2)=\braket{\hat{c}^{\dag}_{j,a} \hat{c}_{j,b}}$ and $\phi(-x/2)=\braket{\hat{c}^{\dag}_{j,a}\hat{c}_{j-1,b}}$ which it is convenient to combine into the even and odd parity hybridizations $\phi_\pm=\phi(x/2)\pm \phi(-x/2)$.   
For later use, we also define $\Delta n = \braket{\hat{n}_{j,b}} - \braket{\hat{n}_{j,a}}$. 
At $t_{ab}=0$, the number of particles on each chain is separately conserved and the model has an associated internal $U(1)$ invariance which is spontaneously broken in the EI state, leading to an one-parameter family of degenerate EI states characterized by $\phi_+=|\phi_+|e^{i\theta_+}\neq 0$ with $\phi_-=0$. 
The collective mode associated with variation of the $U(1)$ phase $\theta_+$ is gapless~\cite{sun2020,murakami2020}. 
When  $t_{ab}\neq 0$, the $U(1)$ symmetry is reduced to $Z_2$, $\phi_-\neq0$ and is real  at all temperatures, and the excitonic order is characterized by the appearance of a nonvanishing $\phi_+$ with the phase $\theta_+= 0$, $\pi$, which spontaneously breaks the $Z_2$ symmetry.  
Because the broken symmetry is discrete, all collective modes are gapped. 
The association of the EI transition with a discrete symmetry breaking is generic in materials~\cite{zenker2014,kaneko2015,murakami2020,mazza2020}; in the EI case considered here the $Z_2$ breaking also makes the $+x$ and $-x$ direction hybridization magnitude different, thereby breaking inversion symmetry  [see Fig.~\ref{fig1}(a)]. 
Following the modern theory of polarization~\cite{king-smith1993,vanderbilt1993,resta1994}, we find the polarization $P=\int \frac{dk}{2\pi} \mathcal{A}_{--}(k) \propto V\phi_+ t_{ab}$, where $\mathcal{A}_{--}(k)$ is the Berry connection at momentum $k$ in the occupied band~\cite{SM}, confirming that when $\phi_+\ne0$ and $t_{ab}\neq 0$ the $Z_2$-broken phase of Eq.~(\ref{H}) is a correlation-driven ferroelectric. 

We solve the model in the time-dependent mean-field (tdMF) theory which captures both the symmetries of the ground state and the needed properties of the collective modes and nonlinear response~\cite{murakami2017,golez2020,bretscher2021_SciAdv,tdMFBSE}. 
The ground state calculation is standard~\cite{SM} and the resulting phase diagram is shown in Fig.~\ref{fig1}(b).  
The dynamics are most conveniently studied in a pseudospin representation $\rho_{\nu}(k,t)\equiv  \braket{\hat{\Psi}_k(t)^{\dag} \sigma_{\nu}\hat{\Psi}_k(t)}/2$ ($\sigma_{\nu}$: Pauli matrix) with $\hat{\Psi}_k^{\dag} =[ \hat{c}^{\dag}_{k,a} , \hat{c}^{\dag}_{k,b}]$.  
We use the equation of motion (EOM) in the length gauge~\cite{sipe2000,SM}: 
\begin{align}
\frac{\partial}{\partial t} \bm{\rho}(k,t) 
=   & 2 \bm{h}(k,t) \times \bm{\rho}(k,t)  
-   E(t) \frac{\partial}{\partial k} \bm{\rho}(k,t) 
\notag \\
&-   \gamma \left[\bm{\rho}(k,t) - \bm{\rho}_{\rm eq}(k)\right],  
\label{eq:EOM}
\end{align}
where $\bm{h}(k,t)$ is the tdMF Hamiltonian in the pseudospin representation 
\begin{align}
h_x(k,t) &= -V {\rm Re}[ \phi_+(t) ] \cos \frac{ k}{2}-  V{\rm Im}[\phi_-(t)] \sin \frac{ k}{2} , 
\notag \\
h_y(k,t) &=(2t_{ab} \!+\! V {\rm Re}[\phi_-(t)]) \sin  \frac{ k}{2}   -V {\rm Im}[ \phi_+(t) ] \cos \frac{ k}{2} ,
\notag \\
h_z(k,t) &= -2t_h\cos k + D + V\Delta n(t).   
\label{eq:mfham}
\end{align} 
Here, we introduce a phenomenological  relaxation term $\gamma$ which may be thought of as the scattering of photoexcited carriers by phonons, disorders, and many-body effects~\cite{attaccalite2013,al-naib2014,mikhailov2016,passos2018,chan2021}.   
At each point in time, the MF parameters are instantaneously updated using the equations $\phi_{\pm} (t) = 2\int \frac{dk}{2\pi}\left[ \rho_x(k,t) + i \rho_y(k,t)\right]\Lambda_{\pm} (k)$ [where $\Lambda_+(k)=\cos\frac{k}{2}$ and $\Lambda_-(k)=i\sin\frac{k}{2}$] and $\Delta n (t) =  -2\int \frac{dk}{2\pi}   \rho_z(k,t)$.  
We solve the equations numerically for a continuous-wave field $E(t) = E_0 \sin \omega_p t$ and initial condition $\bm{\rho}(k,t \! = \! 0) = \bm{\rho}_{\rm eq}(k)$~\cite{SM}. 

\begin{figure}[t]
\begin{center}
\includegraphics[width=\columnwidth]{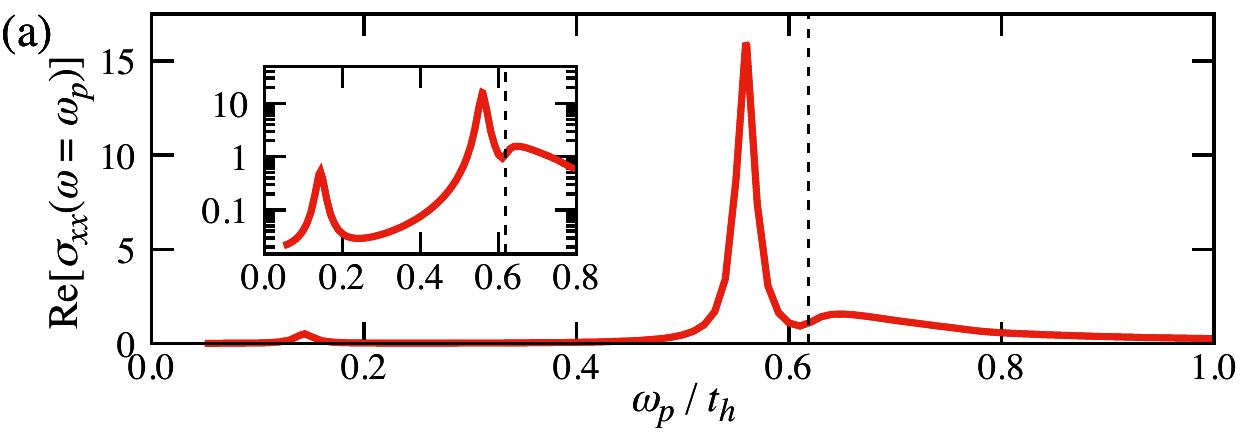}
\\
\includegraphics[width=\columnwidth]{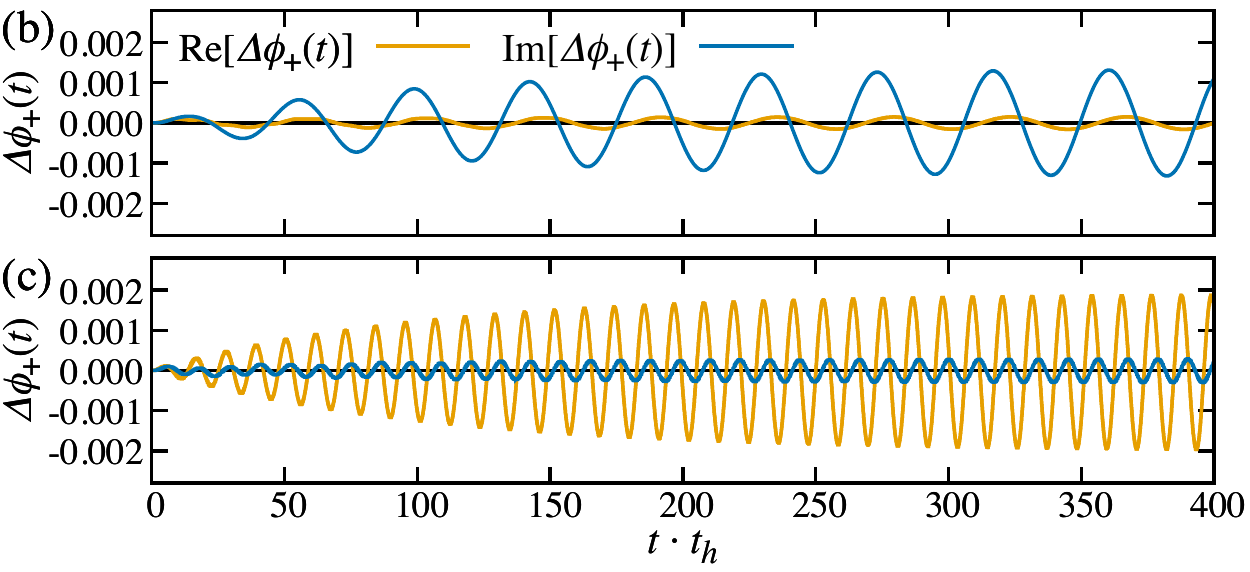}
\caption{(a) Optical conductivity $\sigma_{xx}(\omega=\omega_p)$ of the FEI on linear (main panel) and logarithmic (inset) scales.  
The black dashed line indicates the band gap. 
(b), (c) Time evolution of the real (orange) and imaginary (blue) components of the order parameter $\phi_+(t)$ at (b) $\omega_p/t_h=0.144$ and (c) $\omega_p/t_h=0.559$, which correspond to the phase and amplitude mode frequencies, respectively, where $\Delta \phi_+(t) =  \phi_+(t)-\phi_+(t \!=\! 0)$ is plotted. 
$D/t_h=1$, $t_{ab}/t_h=0.2$, $V/t_h=1.1$, $E_0/t_h = 0.0001$, and $\gamma/t_h=0.01$ are used. }
\label{fig2}
\end{center}
\end{figure}

Figure~\ref{fig2}(a) shows the optical conductivity $\sigma_{xx}(\omega)$ defined here as the Fourier coefficient $J(\omega=\omega_p)$ of the (total) current $J (t) =  2  \int \frac{dk}{2\pi}[\partial_k \bm{h}(k,t)] \cdot \bm{\rho}(k,t)$ at steady state ($t\gg \gamma^{-1}$).  
In the FEI, $\sigma_{xx}(\omega)$ exhibits two peaks below the band gap, arising from the collective modes of the ordered state. 
The collective modes are optically active because for $t_{ab}\ne0$ the ground state breaks inversion symmetry. 
The reduction of the $U(1)$ symmetry down to $Z_2$ by $t_{ab}\neq 0$ means that the two modes are each gapped even at longest wavelength. 
They have mixed phase and amplitude character, but the lower (upper) mode is of dominantly phase (amplitude) mode character.  
As can be seen from Fig.~\ref{fig2}(b), when the system is excited at the frequency of the lower peak, the imaginary part of $\phi_+(t)$ oscillates more strongly than the real part. 
On the other hand,  when the system is excited at the frequency of the upper peak, the real part of $\phi_+(t)$ strongly oscillates [see Fig.~\ref{fig2}(c)].  
Note that the upper peak is separated from the continuum because the band gap including the interchain hopping $t_{ab}$ is larger than the gap originated from $V\phi_+$. 
Since these two modes are optically active, their contributions to higher-order optical responses are important. 

\begin{figure}[t]
\begin{center}
\includegraphics[width=0.965\columnwidth]{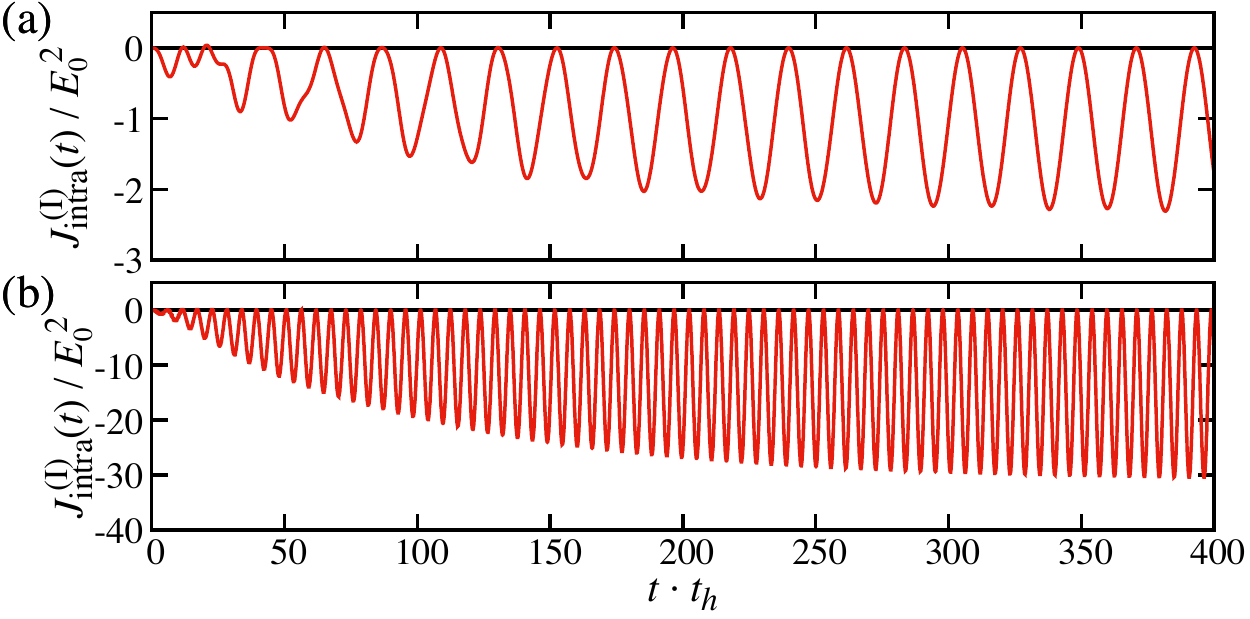}
\\
\includegraphics[width=0.965\columnwidth]{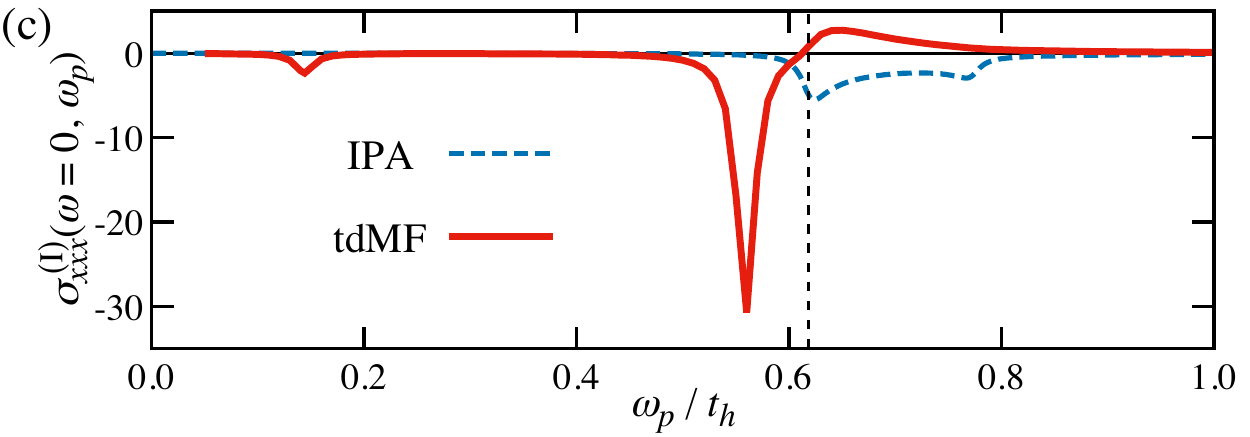}
\caption{(a), (b) Time evolution of the intraband current $J_{\rm intra}^{\rm (I)}(t)$ in the FEI at (a) $\omega_p/t_h=0.144$ and (b) $\omega_p/t_h=0.559$, which correspond to the collective mode frequencies in Figs.~\ref{fig2}(b) and \ref{fig2}(c), respectively. 
(c) Conductivity $\sigma^{\rm (I)}_{xxx}(\omega=0;\omega_p)$ of the shift current. 
The red solid and blue dashed lines indicate $\sigma^{\rm (I)}_{xxx}(\omega=0;\omega_p)$ in the tdMF and IPA, respectively. 
The parameter set is the same as Fig.~\ref{fig2}. 
}
\label{fig3}
\end{center}
\end{figure}

We now discuss the BPVE, which we define as the dc limit of the  intraband current  $J_{\rm intra}$ produced by an applied ac electric field.  
We derive $J_{\rm intra}$ from the time derivative of intraband polarization $P_{\rm intra}$~\cite{sipe2000}. 
Details are given in the Supplemental Material~\cite{SM}. 
We find
\begin{align}
J_{\rm intra}(t)=\int \frac{dk}{2\pi} {\rm tr}\left[\mathcal{J}(k,t)\tilde{\rho}(k,t)\right], 
\label{Jintra}
\end{align}
where $\mathcal{J}$ is defined below and the density matrix $\tilde{\rho}(k,t)$ is obtained via EOM~(\ref{eq:EOM}), which comprises $\tilde{\rho}(k,t)=\tilde{\rho}_{\rm eq}(k)+\tilde{\rho}^{(1)}(k,t)+\tilde{\rho}^{(2)}(k,t)+\cdots$ [superscript indicates order in powers of $E(t)$]. 
The BPVE is second order in $E$ and following prior work we distinguish shift and injection current contributions which are most conveniently written in the band basis (labeled here by $n,m$) of instantaneous eigenstates of $\bm{h}(k,t)$ in Eq.~(\ref{eq:mfham}) [where $\tilde{\rho}(k,t)$ is defined in the band basis]. 

The shift current contribution arises from the generalized derivative $r_{nm;k}(k)= \partial_k \mathcal{A}_{nm}(k) - i\left[\mathcal{A}_{nn}(k) -\mathcal{A}_{mm}(k) \right] \mathcal{A}_{nm}(k)$ ($n\ne m$), which gives the shift vector related to polarization described by the Berry connection~\cite{sipe2000,fregoso2017}, where $\mathcal{A}_{nm}(k) = i\bm{U}^{\dag}_{n}(k) \partial_k \bm{U}_{m}(k)$ is defined by the eigenvector $\bm{U}_n(k)$ of the band $\varepsilon_n(k)$.  
$\mathcal{J}$ of the shift current is given by
\begin{align}
\mathcal{J}^{\rm (I)}_{nm}(k,t) =-E(t) r_{nm;k}(k,t).  
\end{align}
Because $\mathcal{J}^{\rm (I)}\propto E(t)$, the density matrix of first order $\tilde{\rho}^{(1)}(k,t)$ contributes to shift current generation. 
Figures~\ref{fig3}(a) and \ref{fig3}(b) show $J^{({\rm I})}_{\rm intra} (t)=\int \frac{dk}{2\pi} {\rm tr} [\mathcal{J}^{\rm (I)}(k,t)\tilde{\rho}(k,t) ]$   computed for applied electric field equal to  the collective mode frequencies.   
We see immediately that while $J^{({\rm I})}_{\rm intra} (t)$ oscillates, there is a nonzero average, which increases smoothly from zero and saturates, implying a dc photocurrent in the long-time limit.  
Since the BPVE is characterized by $J(\omega=0)=2\sigma_{xxx}(\omega=0;\omega_p) |E(\omega_p)|^2$, it is useful to present the results in terms of the nonlinear conductivity $\sigma_{xxx}(\omega=0;\omega_p) $ defined as  $\frac{2}{E_0^2T_p} \int^{t_m+T_p}_{t_m} J(t)  dt$, where $T_p = 2\pi/\omega_p$ and $t_m$ ($\gg \gamma^{-1}$) is a time long enough for steady state to be reached~\cite{SM}. 
Figure~\ref{fig3}(c) shows $\sigma^{\rm (I)}_{xxx}(\omega=0;\omega_p)$ corresponding to our results for $J^{({\rm I})}_{\rm intra} (t)$.  
Corresponding to Figs.~\ref{fig3}(a) and \ref{fig3}(b), $\sigma^{\rm (I)}_{xxx}(0;\omega_p)$ in the tdMF shows two sharp peaks in the sub-band-gap regime, indicating that excitation of collective modes (especially the amplitudelike mode) makes a large contribution to the shift current.  
For comparison, we also plot  in Fig.~\ref{fig3}(c) $\sigma^{\rm (I)}_{xxx}(0;\omega_p)$ obtained from the independent particle approximation (IPA) that assumes that the MF parameter magnitude and phase remain fixed during the excitation process. 
We see that in the IPA the conductivity is nonzero only in the above-band-gap regime and is smaller in magnitude than the amplitude mode contribution, showing that the collective dynamics of the electronically ordered state make a large contribution to the nonlinear conductivity.

\begin{figure}[t]
\begin{center}
\includegraphics[width=\columnwidth]{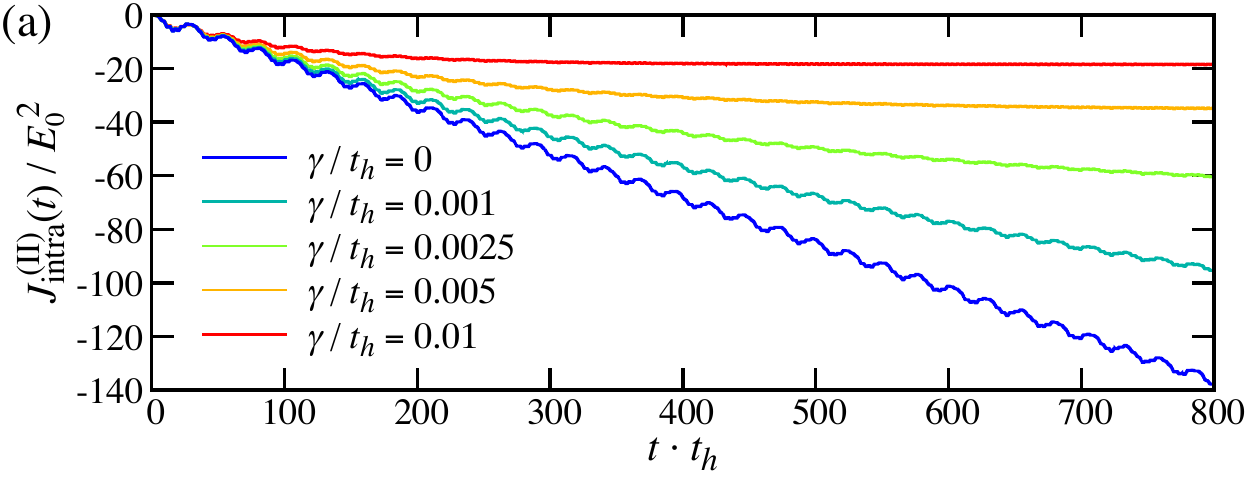}
\\
\includegraphics[width=\columnwidth]{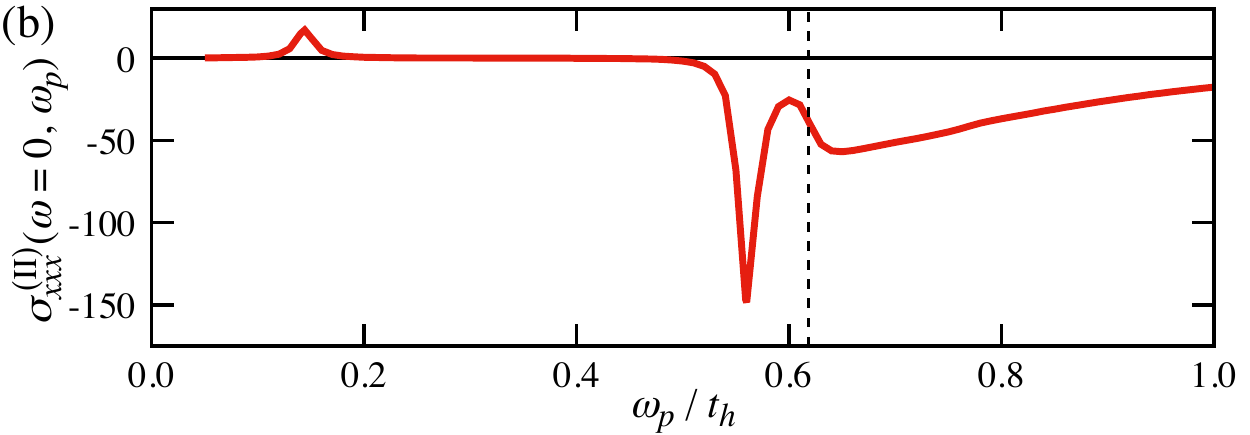}
\caption{(a) Time evolution of the intraband current $J_{\rm intra}^{\rm (II)}(t)$ in the FEI at $\omega_p/t_h=0.8$. 
(b) Conductivity $\sigma_{xxx}^{\rm (II)}(\omega=0;\omega_p)$ of the injection current  with $\gamma/t_h=0.01$.  
$D/t_h=1$, $t_{ab}/t_h=0.2$, $V/t_h=1.1$, and $E_0/t_h = 0.0001$ are used. 
}
\label{fig4}
\end{center}
\end{figure}

\begin{figure}[t]
\begin{center}
\includegraphics[width=\columnwidth]{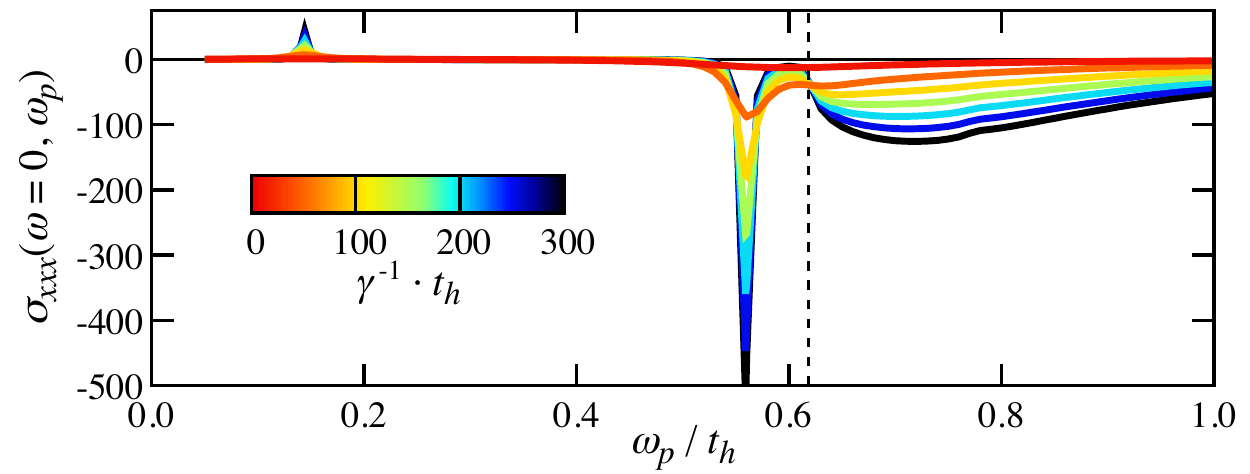}
\caption{
Conductivity $\sigma_{xxx}(\omega=0;\omega_p)=\sigma_{xxx}^{\rm (I)}(\omega=0;\omega_p)+\sigma_{xxx}^{\rm (II)}(\omega=0;\omega_p)$ with changing $\gamma$.  
The parameter set is the same as Fig.~\ref{fig4}. 
}
\label{fig5}
\end{center}
\end{figure}

Next, we examine the  injection current contribution~\cite{sipe2000},  for which 
\begin{align}
\mathcal{J}^{\rm (II)}_{nm}(k,t)= v_{n}(k,t) \delta_{nm},
\end{align}
where $v_{n}(k,t)= \partial_k \varepsilon_{n} (k,t)$. 
Because $\mathcal{J}^{\rm (II)}$ does not contain $E(t)$, the density matrix of second order $\tilde{\rho}^{(2)}(k,t)$ contributes to injection current generation. 
Figure~\ref{fig4}(a) shows the injection current contribution to the BPVE for different values of the phenomenological relaxation $\gamma$. 
When $\gamma \ne 0$,  $J^{({\rm II})}_{\rm intra} (t)$ increases linearly at short times and saturates for times $\gg \gamma^{-1}$, which is a characteristic of the injection current~\cite{zhang2019,holder2020}. 
Our result for the FEI is contrasted to the IPA result that for a free system with time-reversal symmetry linearly polarized light does not produce an injection current. 
In Fig.~\ref{fig4}(b), we plot the nonlinear conductivity $\sigma^{({\rm II})}_{xxx}(\omega=0;\omega_p)$ corresponding to $J^{({\rm II})}_{\rm intra} (t)$, which exhibits two sharp peaks at the sub-band-gap collective mode frequencies and in addition the large broadband contribution in the above-band-gap regime.     
Since the injection current is of order $\gamma^{-1}$~\cite{zhang2019,SM}, it can correspond to a strong optical response when dissipative effects are small.  
As shown in Fig.~\ref{fig5}, the conductivity $\sigma_{xxx}^{\rm (I)}(0;\omega_p)+\sigma_{xxx}^{\rm (II)}(0;\omega_p)$ exhibits large values with increasing $\gamma^{-1}$ due to the injection current contribution.  

Finally, we show how the  injection current arises from photoinduced deformations of the order parameter (see the Supplemental Material~\cite{SM} for details). 
In the velocity gauge~\cite{parker2019,holder2020}, the injection current may be written  
\begin{align}
J_{\rm IC}(0;\omega_p)  
= \! \int \! &\frac{dk}{2\pi} \! \int \! \! \frac{d\omega'}{2\pi} \, 
{\rm tr} \Bigl[ \mathcal{V}_0(k) G_0(k,\omega') \delta \Delta^{(1)}(k,-\omega_p) 
\notag \\
&\;\; \times G_0(k,\omega' +\omega_p) \delta\Delta^{(1)}(k,\omega_p)  G_0(k,\omega') \Bigr]  
\notag 
\\ &+ [\omega_p \, \leftrightarrow \, -\omega_p], 
\label{eq:JIC}
\end{align}
where $G_0(k,\omega)$ is the bare fermion propagator and $\mathcal{V}_0(k)=t_{ab}\cos (k/2) \sigma_2 + 2t_h\sin(k) \sigma_3$ is the $k$ derivative of the tight-binding Hamiltonian. 
$\delta \Delta^{(1)}(k,\omega_p)$ is the perturbation at first order of $A(\omega_p) \propto E(\omega_p)/\omega_p$ (vector potential). 
If the MF parameters were fixed to their equilibrium values (i.e., IPA), $\delta \Delta^{(1)}(k,\omega_p) = -\mathcal{V}_0(k) A(\omega_p)$ and the injection current in Eq.~(\ref{eq:JIC}) vanishes due to $\mathcal{V}_0(-k)=-\mathcal{V}_0(k)^*$~\cite{SM}. 
However, in the FEI the incident light modulates $\delta\Delta_{\pm}(t) =V\phi_{\pm}(t)-V\phi^{\rm eq}_{\pm}$ and $\delta n(t)=\Delta n(t)-\Delta n^{\rm eq}$ [see e.g., Fig.~\ref{fig2}] so that the total perturbation $\delta \Delta^{(1)}(k,\omega_p)=-\mathcal{V}_0(k) A(\omega_p)+\delta \Delta^{(1)}_{\rm MF}(k,\omega_p)$ contains $\delta \Delta^{(1)}_{\rm MF}(k,\omega_p)=\delta\bm{\Delta}(k,\omega_p)\cdot \bm{\sigma}$ given by  
\begin{align}
\delta \Delta_{x}(k,  \omega_p) 
& =  -\delta \Delta^{\rm \! R}_+ (\omega_p) \cos  \frac{k}{2} - \delta \Delta^{\rm I}_- (\omega_p) \sin  \frac{k}{2}  ,
\notag \\
\delta \Delta_y(k,  \omega_p) 
&  =  - \delta \Delta^{\rm I}_+  (\omega_p) \cos \! \frac{k}{2}  + \delta \Delta^{\rm \! R}_- (\omega_p)  \sin  \frac{k}{2},
\notag \\
\delta \Delta_z(k, \omega_p) 
&  =  V \delta n(\omega_p) ,
\label{eq:pertub}
\end{align}
where the superscripts ${\rm R}$ and ${\rm I}$ indicate the real and imaginary part of the order parameter, respectively. 
The inversion-breaking nature of the FEI phase means $\delta \Delta_{\pm}(\omega_p) \propto A(\omega_p)$ and $\delta \Delta^{(1)}(-k,-\omega_p) \ne - \delta \Delta^{(1)}(k,\omega_p)^*$, which makes a nonvanishing contribution to the integrands in Eq.~(\ref{eq:JIC}). 
Hence, order parameter deformations produce a nonvanishing injection current.      

In summary, we have investigated the shift and injection current contributions to the BPVE  in a correlated inversion-symmetry-breaking insulator: the FEI. 
The physics of the correlated insulator produces characteristic enhancements of the BPVE, related to the deformability of the order parameter under applied electric fields.   
The shift current is modified and shows sharp resonances at the collective mode frequencies. 
The injection current has both resonant contributions at the collective mode frequencies and a broadband contribution at above-band-gap drive frequencies which arises from the deformability of the order parameter and is entirely absent in a rigid band picture. 
It is very weak in the phonon-driven ferroelectric case because the energy scale mismatch between the phonon frequency and electronic band gap weakens the influence of the phonon motion on the electronic system~\cite{SM}.  

In contrast to the previous studies that address excitonic effects~\cite{,morimoto2016ex,fei2020,morimoto2020,chan2021}, we focus on collective order-parameter dynamics in a symmetry-broken state and reveal its effects on the BPVE.  
While we used a simple one-dimensional model of a FEI, our idea is applicable to higher dimensions and richer models. 
The essential ingredient is  a deformable electronic order parameter that produces a broken symmetry. 
Although our two-chain model is similar to models proposed for Ta$_2$NiSe$_5$~\cite{mazza2020}, in Ta$_2$NiSe$_5$ the ordered state is inversion symmetric and the BPVE vanishes. 
However, under an applied bias voltage, Ta$_2$NiSe$_5$ has shown the light-intensity-dependent photocurrent generation~\cite{li2016}. 
The mechanism of current generation under both dc and optical electric fields is an interesting open question.  The FEI has recently been predicted in the monolayer transition metal dichalcogenides~\cite{varsano2020}, which is a potential candidate that exhibits the BPVE.  
The search for materials candidates should include the properties (a) inversion-symmetry breaking [e.g., the ferroelectric state] and (b) strong electronic character of the order. 
\\


This work was supported by Grants-in-Aid for Scientific Research from JSPS, KAKENHI Grants No.~JP18K13509 (T.K.), No.~JP19K23425, N o.~JP20K14412, No.~JP20H05265 (Y.M.) and JST CREST Grant No.~JPMJCR1901 (Y.M.). 
A.J.M. and Z.S. acknowledge support from the Energy Frontier Research Center on Programmable Quantum Materials funded by the U.S. Department of Energy (DOE), Office of Science, Basic Energy Sciences (BES), under Award No.~DE-SC0019443.
The numerical calculations were performed in part using computational resources at RIKEN. 
T.K. was supported by the JSPS Overseas Research Fellowship. 
D.G. is supported by the Slovenian Research Agency (ARRS) under Program  J1-2455 and P1-0044. 
The Flatiron Institute is a division of the Simons Foundation.


\bibliography{References_arXiv}

\begin{thebibliography}{82}
\bibitem [{ceh()}]{ceha}%
  \BibitemOpen
  \href@noop {} {}\bibinfo {note} {The light velocity ($c$), the charge ($q$),
  the Planck constant ($\hbar$), and the lattice constant ($a$) are set to 1.
  In Figs.~3--5, $\sigma_{xxx}(\omega=0;\omega_p)$ [$J_{\rm intra}(t)/E_0^2$]
  is plotted in units of $q^3a^2/ \hbar t_h \propto t_h^{-1}$.}\BibitemShut
  {Stop}%
\bibitem [{tdM()}]{tdMFBSE}%
  \BibitemOpen
  \href@noop {} {}\bibinfo {note} {The tdMF calculation is essentially
  equivalent to the solution of the Bethe-Salpeter equation with the vertex
  correction.}\BibitemShut {Stop}%
\end{thebibliography}

\clearpage

\renewcommand\thesection{}
\renewcommand\thefigure{S.\arabic{figure}}
\setcounter{figure}{0}
\renewcommand{\bibnumfmt}[1]{[S#1]}
\renewcommand{\citenumfont}[1]{S#1}

\subsection*{\normalsize Supplemental Material:\\
Bulk Photovoltaic Effect Driven by Collective Excitations in a Correlated Insulator}
\vspace{3pt}

\begin{appendix}


\section{Hartree-Fock Approximation}

\subsection{Hamiltonian} 
To study the excitonic insulator (EI) state, we employ the Hartree-Fock (HF) approximation for the interaction term $\hat{\mathcal{H}}_{V}$.  
In the HF approximation, we have the Hamiltonian $\hat{\mathcal{H}}^{\rm HF}_{V}=\hat{\mathcal{H}}^{\rm H}_{V} + \hat{\mathcal{H}}^{\rm F}_{V} +  N\varepsilon^{\rm HF}_V$ with 
\begin{align}
\hat{\mathcal{H}}^{\rm H}_{V} =& 
V\left(n + \Delta n\right) \sum_{j}   \hat{n}_{j,a} + V\left(n - \Delta n\right)  \sum_{j}   \hat{n}_{j,b}, 
\\
\hat{\mathcal{H}}^{\rm F}_{V} = 
&-V  \sum_{j}    \left( \phi \left(+\frac{x}{2}\right) \hat{c}^{\dag}_{j,b} \hat{c}_{j,a} + {\rm H.c.}\right) 
\notag \\
&-V \sum_{j}   \left(  \phi \left(-\frac{x}{2}\right) \hat{c}^{\dag}_{j-1,b} \hat{c}_{j,a} + {\rm H.c.} \right), 
\end{align}
where we assumed $n = n_a + n_b$ and $\Delta n = n_b - n_a$ with $n_{\alpha} = \braket{\hat{n}_{j,\alpha}} =\braket{\hat{n}_{j-1,\alpha}}$ ($\alpha$=$a, b$) in the Hartree term $\hat{\mathcal{H}}^{\rm H}_{V}$ and $\phi(+x/2) = \braket{\hat{c}^{\dag}_{j,a} \hat{c}_{j,b}}$ and $\phi(-x/2) = \braket{\hat{c}^{\dag}_{j,a} \hat{c}_{j-1,b}}$ in the Fock term $\hat{\mathcal{H}}^{\rm F}_{V}$.  

Due to the interchain hopping $t_{ab}$, $\phi (+x/2)=-\phi (-x/2)$ without the excitonic order. 
This symmetry is broken when the excitonic order parameter 
\begin{align}
\phi_+ =  \phi\left(+\frac{x}{2}\right)  + \phi \left(-\frac{x}{2}\right)  ,
\end{align}
is nonzero~\citeSM{mazza2020SM}. 
Hence, we define 
\begin{align}
&\phi\left(+\frac{x}{2}\right)  =  \frac{\phi_+}{2} + \frac{\phi_-}{2},
\notag \\
&\phi \left(-\frac{x}{2}\right)  = \frac{\phi_+}{2} - \frac{\phi_-}{2} .
\end{align} 
Using the Fourier transformation
\begin{align}
\hat{c}_{j,a} &= \frac{1}{\sqrt{N}} \sum_{k} e^{ikR_j} \hat{c}_{k,a}, 
\notag \\
\hat{c}_{j,b} &= \frac{1}{\sqrt{N}} \sum_{k} e^{ik(R_j+1/2)} \hat{c}_{k,b} ,
\end{align}
and  $\hat{\Psi}_k^{\dag} =[ \hat{c}^{\dag}_{k,a} , \hat{c}^{\dag}_{k,b} ]$, we obtain the HF Hamiltonian 
$\hat{\mathcal{H}}^{\rm HF} = \sum_k \hat{\Psi}_k^{\dag}\mathcal{H}_k \hat{\Psi}_k + Vn\sum_{k} \hat{\Psi}_k^{\dag}\hat{\Psi}_k+ N\varepsilon^{\rm HF}_V$ with 
\begin{widetext}
\begin{align}
\mathcal{H}_k = 
\left[
\begin{array}{cc}
-2t_h\cos k +D + V\Delta n & -V\phi_+^* \cos \frac{k}{2} -i(2t_{ab} + V\phi_-^*) \sin\frac{k}{2}
\\
 -V\phi_+ \cos \frac{k}{2} +i(2t_{ab} + V\phi_-) \sin \frac{k}{2}  &  2t_h\cos k - D - V\Delta n 
\end{array}
\right] , 
\notag 
\end{align}
\end{widetext}
where $t_a = -t_b = t_h$ and the lattice constant is set to 1. 
The energy term $\varepsilon^{\rm HF}_V$ is given by
\begin{align}
\varepsilon^{\rm HF}_V= -\frac{V}{2} \left( n^2 - \Delta n ^2 \right) + \frac{V}{2} \left( |\phi_+ |^2 + |\phi_-|^2 \right).
\end{align}
When the hybridization parameters are complex (e.g., $\phi_+ = {\rm Re}[\phi_+] + i {\rm Im}[\phi_+]$), the Hamiltonian $\mathcal{H}_k = \bm{h}(k)\cdot \bm{\sigma}$  in the pseudospin representation is given by 
\begin{align}
h_x(k) &= -V {\rm Re}[ \phi_+] \cos \frac{ k}{2}-  V{\rm Im}[\phi_-] \sin \frac{ k}{2} , 
\label{eq:hameq_x}\\
h_y(k) &=(2t_{ab} + V {\rm Re}[\phi_-]) \sin  \frac{ k}{2}   -V {\rm Im}[ \phi_+ ] \cos \frac{ k}{2} ,
\label{eq:hameq_y}\\
h_z(k) &= -2t_h\cos k + D + V\Delta n. 
\label{eq:hameq_z}
\end{align}

\subsection{Self-consistent equations}
The Hamiltonian $\mathcal{H}_k = \bm{h}(k)\cdot \bm{\sigma}$ has the eigenenergies 
\begin{align}
\varepsilon_{\pm}(k) = \pm |\bm{h}(k)| = \pm \sqrt{ h_x(k)^2 + h_y(k)^2 +  h_z(k)^2}.  
\end{align}
The eigenvectors of the upper ($+$) and lower ($-$) energy bands are given by
\begin{align}
&\bm{U}_+(k) =
\frac{1}{\sqrt{2}}
\left[
\begin{array}{c}
u_k 
\\
v_ke^{i\varphi_k}
\end{array}
\right]  ,
\label{eq:eigenstates_1} \\
& \bm{U}_-(k) = 
\frac{1}{\sqrt{2}}
\left[
\begin{array}{c}
v_k 
\\
-u_ke^{i\varphi_k}
\end{array}
\right], 
\label{eq:eigenstates_2}
\end{align}
respectively, with 
\begin{align}
&u_k  = \sqrt{1+\frac{h_z(k)}{ |\bm{h}(k)| }},
\label{eq:eigenstates_uk}  \\ 
&v_k = \sqrt{1-\frac{h_z(k)}{ |\bm{h}(k)| }},
\label{eq:eigenstates_vk} \\ 
&\varphi_k  = \tan^{-1}\left[ \frac{h_y(k)}{h_x(k)} \right] . 
\label{eq:eigenstates_par}
\end{align} 

Using the eigenvectors, the total particle density $n$ and the difference of occupancy $\Delta n$ are given by 
\begin{align}
n &=  \frac{1}{N} \sum_{k} \sum_{n=\pm} f(\varepsilon_n(k)),  
\\
\Delta n &=  -\frac{1}{N} \sum_{k}  \frac{h_z(k)}{ |\bm{h}(k)| }  f_{+-}(k),  
\label{eq:eqdn}
\end{align}
respectively. 
Here, $f_{+-}(k) = f(\varepsilon_+(k)) - f(\varepsilon_-(k))$ is the difference of the Fermi distribution function $f(\varepsilon)$,  and $f_{+-}(k)=-1$ at zero temperature. 
The hybridization parameters $\phi_+$ and $\phi_- $ are given by 
\begin{align}
\phi_+ &  =    \frac{1}{N} \! \sum_{k}  \frac{h_x(k) + i h_y(k)}{|\bm{h}(k)|} f_{+-}(k) \cos  \frac{ k}{2}  ,
\label{eq:eqphip}
\\
\phi_-  &  =   \frac{i}{N} \! \sum_{k} \frac{ h_x(k) +i h_y(k)}{|\bm{h}(k)|} f_{+-}(k) \sin  \frac{ k}{2}  ,
\label{eq:eqphim}
\end{align} 
respectively. 
These equations correspond to the gap equations and we solve them self-consistently.


\section{Equation of Motion}
Here, we derive the equation of motion (EOM) for the density 
\begin{align}
\rho_{\beta\alpha}(k,t) =  \braket{ \hat{c}^{\dag}_{k,\alpha}(t) \hat{c}_{k,\beta}(t)  }
\end{align}
in the orbital basis. 
In this section, we also use the density operator defined by $\hat{\rho}_{\beta\alpha,k} =  \hat{c}^{\dag}_{k,\alpha} \hat{c}_{k,\beta}$. 

Employing the length gauge~\citeSM{sipe2000SM}, we consider the time-dependent Hamiltonian under the electric field $E(t)$ given by
\begin{align}
\hat{\mathcal{H}}(t) = \hat{\mathcal{H}}_0  -E(t)  \hat{P}. 
\label{eq:tdHamEP}
\end{align}
$\hat{\mathcal{H}}_0$ is the single-particle term 
\begin{align}
\hat{\mathcal{H}}_0  = \sum_{k} \sum_{\alpha,\beta} h_{\alpha\beta}(k) \hat{c}^{\dag}_{k,\alpha}  \hat{c}_{k,\beta}. 
\label{eq:tdHam0}
\end{align}
In our tdMF theory, this term corresponds to the HF Hamiltonian $\hat{\mathcal{H}}^{\rm HF}$.  
$\hat{P}$ in Eq.~(\ref{eq:tdHamEP}) is the polarization defined as 
\begin{align}
\hat{P} &= \sum_{j,\alpha} 
R_{j,\alpha}\hat{c}^{\dag}_{j,\alpha}\hat{c}_{j,\alpha}
\\
&=  \sum_{k,k'}\sum_{\alpha} \left[  i \frac{\partial}{\partial k}  \delta(k-k') \right] \hat{c}^{\dag}_{k,\alpha}  \hat{c}_{k',\alpha} ,
\label{Porbital}
\end{align} 
where $R_{j,\alpha} = R_j+r_\alpha$ and $R_j$ and $r_\alpha$ are the positions of the $j$-th unit cell and atomic orbital $\alpha$, respectively. 
Here, we set $\hbar=q=1$ for simplicity.    
Because we are considering two different orbitals on the different sites in the tight-binding picture, we assume $\bm{d}_{\alpha\beta} = \int d\bm{r} w_{\alpha}(\bm{r}-\bm{r}_{\alpha}) \bm{r} w_{\beta}(\bm{r}-\bm{r}_{\beta})=0$ ($\bm{r}_{\alpha}\ne \bm{r}_{\beta}$). 

We derive the EOM from the Heisenberg EOM 
\begin{align}
&i\frac{\partial}{\partial t}\hat{\rho}_{\beta\alpha,k}(t) =  \left[ \hat{\rho}_{\beta\alpha,k}(t) , \hat{\mathcal{H}}(t) \right] . 
\end{align}
Because 
\begin{align}
 \left[ \hat{\rho}_{\beta\alpha,k} , \hat{\mathcal{H}}_0 \right] = \sum_{\gamma} \left[ h_{\beta\gamma} (k) \hat{\rho}_{ \gamma \alpha,k} - \hat{\rho}_{\beta \gamma,k} h_{\gamma\alpha} (k)  \right] 
\end{align}
and 
\begin{align}
&\left[ \hat{\rho}_{\beta \alpha,k} , \hat{P} \right] = i \frac{\partial}{\partial k}\hat{\rho}_{\beta \alpha,k}  , 
\end{align}
the EOM for $\rho_{\beta\alpha}(k,t)=\braket{\hat{\rho}_{\beta\alpha,k}(t)}$ is given by 
\begin{align}
\frac{\partial}{\partial t} \rho_{\beta\alpha}(k,t) 
= & - i \sum_{\gamma} \left[ h_{\beta\gamma} (k) \rho_{\gamma \alpha}(k,t) - \rho_{\beta \gamma} (k,t)h_{\gamma\alpha} (k)  \right]
\notag \\
&-  E(t) \frac{\partial}{\partial k} \rho_{\beta \alpha}  (k,t). 
\end{align}
Using the pseudospin representation 
\begin{align}
&\left[
\begin{array}{cc}
\rho_{aa}(k,t) & \rho_{ab}(k,t)
\\
\rho_{ba}(k,t) & \rho_{bb}(k,t)
\end{array}
\right]
=\rho_0(k,t) \sigma_0 + \bm{\rho}(k,t) \cdot \bm{\sigma}, 
\end{align}
the EOM for $\bm{\rho}(k,t)$ is given by
\begin{align}
&\frac{\partial}{\partial t} \bm{\rho}(k,t) 
=    2\bm{h}(k) \times \bm{\rho}(k,t)  
 -  E(t)\frac{\partial}{\partial k} \bm{\rho}(k,t) . 
\end{align}

The MF parameters 
\begin{align}
\phi_+ (t)&  =    \frac{2}{N} \! \sum_{k}  \left[ \rho_x(k,t) + i \rho_y(k,t)\right] \cos  \frac{ k}{2}  ,
\label{eq:tdphip}
\\
\phi_-(t)  &  =   \frac{2i}{N} \! \sum_{k} \left[ \rho_x(k,t) + i \rho_y(k,t)\right] \sin  \frac{ k}{2}  ,
\label{eq:tdphim}
\end{align} 
and 
\begin{align}
\Delta n (t) &=  -\frac{2}{N} \sum_{k}  \rho_z(k,t) 
\label{eq:tddn}
\end{align}
are also updated in time. 
In the tdMF, $\bm{h}(k)\rightarrow \bm{h}(k,t) = \bm{h}^{\rm MF}[\phi_{\pm},\Delta n](k,t)$,  and the EOM is given by
\begin{align}
&\frac{\partial}{\partial t} \bm{\rho}(k,t) 
=  2\bm{h}(k,t) \times \bm{\rho}(k,t)  
 -  E(t)\frac{\partial}{\partial k} \bm{\rho}(k,t) . 
\end{align}

In equilibrium, the particle density  is given by 
\begin{align}
\bm{\rho}_{\rm eq} (k) =  \bm{\rho}(k,t=0)  = \frac{\bm{h}(k)}{2|\bm{h}(k)|} f_{+-} (k). 
\end{align}
In the time-dependent calculations, we employ the ground state in equilibrium as the initial state and use the Runge-Kutta fourth-order method for time evolution. 
In our calculation, we use the time step $\delta t = T_p / N_t $ with $T_p = 2\pi / \omega_p$ and $N_t=500$. 
For $k$ derivative, we use the numerical differentiation $\partial_k \bm{\rho}(k,t) = [\bm{\rho}(k+dk,t) - \bm{\rho}(k-dk,t)]/(2dk)$ (symmetric derivative), where $dk=2\pi/N$ with $N>5000$.


\section{Current}

\subsection{Current in the orbital basis} 
The current operator is defined as 
\begin{align}
\hat{J} (t) = \frac{d\hat{P}(t)}{dt} = \frac{1}{i} \left[ \hat{P} (t), \hat{\mathcal{H}}(t) \right].
\end{align}
In the length gauge, $[ \hat{P} , \hat{\mathcal{H}}_0 - E(t) \hat{P} ] = [ \hat{P} , \hat{\mathcal{H}}_0 ] $. 
Using the Hamiltonian (\ref{eq:tdHam0}),  we have 
\begin{align}
\left[ \hat{P} , \hat{\mathcal{H}}_0\right] 
=  i \sum_{k} \sum_{\alpha,\beta}  \frac{\partial h_{\alpha\beta}(k)}{\partial k} \hat{\rho}_{\beta\alpha,k}.
\end{align}
Hence, the current per unit [$J(t) = \braket{\hat{J}(t)}/N$] is given by 
\begin{align}
J (t) =  \int \frac{dk}{2\pi}\sum_{\alpha,\beta}  \frac{\partial h_{\alpha\beta}(k,t)}{\partial k} \rho_{\beta\alpha}(k,t). 
\end{align} 
Using the pseudospin representation in the two-orbital system, the current is written as   
\begin{align}
J (t) =  2\int \frac{dk}{2\pi} \, \frac{\partial \bm{h}(k,t)}{\partial k} \cdot \bm{\rho}(k,t), 
\label{eq:jtot_orb}
\end{align}
where $h_{aa}(k,t) + h_{bb}(k,t)=0$ (particle-hole symmetry) is assumed. 
Note that, in our tdMF calculation, we obtain the same value of $J(t)$ even if the MF parameters are not included in $\partial_k \bm{h}(k,t)$.

\subsection{Intraband current} 
The current operator derived in the previous subsection includes both inter and intraband contributions. 
The dc current needed in the bulk photovoltaic effect (BPVE) arises from the intraband contribution to the current~\citeSM{sipe2000SM}.  
To derive this contribution and make contact with previous work~\citeSM{sipe2000SM}, we transform  the polarization and current  operators from the orbital basis $\hat{c}_{k\alpha}$ to the band basis $\hat{d}_{k,n}$  that diagonalizes $\hat{\mathcal{H}}^{\rm HF}$:
\begin{align}
\sum_{\alpha,\beta} h_{\alpha\beta}(k) \hat{c}^{\dag}_{k,\alpha}  \hat{c}_{k,\beta}
= 
\sum_{n} \varepsilon_{n}(k) \hat{d}^{\dag}_{k,n}  \hat{d}_{k,n}.
\end{align}
The unitary transformation that gives the relation between the operators of the orbital $\alpha$ and band $n$ is 
\begin{align}
\hat{c}_{k,\alpha} = \sum_{n} U_{\alpha n} (k)  \hat{d}_{k,n}. 
\label{transformation}
\end{align} 
For later use, we define the density operator in the band basis 
\begin{align}
\hat{\tilde{\rho}}_{mn,k} =  \hat{d}^{\dag}_{k,n} \hat{d}_{k,m}, 
\end{align}
where we put $\tilde{}$ on $\rho$ to indicate the band basis.  
We also use its expectation value $\tilde{\rho}_{mn}(k,t) =  \braket{ \hat{d}^{\dag}_{k,n}(t) \hat{d}_{k,m}(t)}$. 

Using Eq.~(\ref{transformation}) in the $k$-representation of Eq.~(\ref{Porbital}), we obtain~\citeSM{sipe2000SM}
\begin{align}
\hat{P} &=\sum_k\sum_{n, m } \mathcal{A}_{nm}(k)  \hat{d}^{\dag}_{k,n} \hat{d}_{k,m}
\notag \\
& +\sum_{k,k'} \sum_{n} \left[  i \frac{\partial}{\partial k}  \delta(k-k') \right] \hat{d}^{\dag}_{k,n} \hat{d}_{k',n}, 
\end{align}
where $\mathcal{A}_{nm}(k)$ is the connection defined as
\begin{align}
\mathcal{A}_{nm}(k) =  i \sum_{\alpha} U^*_{\alpha n}(k) \frac{\partial}{\partial k}  U_{\alpha m}(k) . 
\end{align}
In this Supplemental Material, following the notation in Ref.~\citeSM{sipe2000SM}, we define
\begin{align}
r_{nm}(k) = 
\begin{cases}
\mathcal{A}_{nm}(k) & (n\ne m )
\\
0  & (n=m )
\end{cases}
,
\end{align}
and divide the polarization into the inter and intraband contributions,
\begin{align}
&\hat{P} = \hat{P}_{\rm inter}  + \hat{P}_{\rm intra}
\end{align}
with  
\begin{align}
\hat{P}_{\rm inter} &= \sum_k \sum_{n \neq m }r_{nm}(k) 
\hat{d}^{\dag}_{k,n} \hat{d}_{k,m}, 
\\
\hat{P}_{\rm intra} &= \sum_k \sum_{n} \mathcal{A}_{nn}(k) \hat{d}^{\dag}_{k,n} \hat{d}_{k,n}
\notag \\
&+\sum_{k,k'} \sum_{n} \left[ i \frac{\partial}{\partial k}  \delta(k-k') \right] \hat{d}^{\dag}_{k,n} \hat{d}_{k',n} . 
\label{eq:Pintra}
\end{align}

Because the shift and injection currents are obtained from the intraband current~\citeSM{sipe2000SM}, we consider 
\begin{equation}
\hat{J}_{\rm intra} (t) = \frac{d\hat{P}_{\rm intra} (t)}{dt} = \frac{1}{i} \left[ \hat{P}_{\rm intra} (t), \hat{\mathcal{H}}(t) \right]
\label{Jintra}
\end{equation}
in the length gauge, where $\hat{\mathcal{H}}(t) = \hat{\mathcal{H}}_0-E(t)(\hat{P}_{\rm intra} +\hat{P}_{\rm inter})$, and  we have 
\begin{align}
\hat{J}_{\rm intra}(t)=\left[ \hat{P}_{\rm intra}, \hat{\mathcal{H}_0}\right]-E(t)\left[\hat{P}_{\rm intra} , \hat{P}_{\rm inter}\right]. 
\end{align} 
Using the Hamiltonian $\hat{\mathcal{H}}_0  = \sum_{k,n} \varepsilon_{n}(k) \hat{d}^{\dag}_{k,n}  \hat{d}_{k,n}$, we have 
\begin{align}
\left[ \hat{P}_{\rm intra} , \hat{\mathcal{H}}_0\right] 
=  i \sum_k \sum_n v_{n}(k) \hat{\tilde{\rho}}_{nn,k}, 
\end{align}
where $v_{n}(k)=\partial_k \varepsilon_n(k)$. 
The commutation relation between the intra and interband polarizations is given by 
\begin{align}
\left[ \hat{P}_{\rm intra}, \hat{P}_{\rm inter} \right] 
&= i \sum_k  \sum_{n\neq m} r_{nm;k}(k) \hat{\tilde{\rho}}_{mn,k} 
\end{align}
with the generalized derivative 
\begin{align}
r_{nm;k}(k) =  \frac{\partial}{\partial k} r_{nm}(k)-i\left[\mathcal{A}_{nn}(k) -\mathcal{A}_{mm}(k) \right] r_{nm}(k). 
\end{align} 
Hence, the intraband current is given by~\citeSM{sipe2000SM}
\begin{align}
\hat{J}_{\rm intra} (t) &=   \sum_k\sum_{n}  v_{n}(k) \hat{\tilde{\rho}}_{nn,k}(t)
\notag \\
&  - E(t)  \sum_k \sum_{n\neq m}  r_{nm;k}(k) \hat{\tilde{\rho}}_{mn,k}(t). 
\label{eq:intraop_ap}
\end{align}
Within the second-order perturbation theory, the first and second terms in Eq.~(\ref{eq:intraop_ap})  are the injection and shift currents, respectively~\citeSM{sipe2000SM}.  
In the tdMF theory, the band parameters are also time-dependent, and we thus define the shift and injection currents (per unit) as 
\begin{align}
&J^{({\rm I})}_{\rm intra} (t) = - E(t) \int  \frac{dk}{2\pi} \sum_{n \ne m} r_{nm;k}(k,t) \tilde{\rho}_{mn}(k,t), 
\label{eq:Jintra1_ap}
\\
&J^{(\rm{II})}_{\rm intra} (t) =   \int \frac{dk}{2\pi} \sum_{n}  v_{n}(k,t) \tilde{\rho}_{nn}(k,t) , 
\label{eq:Jintra2_ap}
\end{align}
respectively. 
Corresponding to them, in the main text, we defined $\mathcal{J}^{\rm (I)}_{nm}(k,t) =-E(t) r_{nm;k}(k,t)$ $(n\ne m)$ and $\mathcal{J}^{\rm (II)}_{nm}(k,t)= v_{n}(k,t) \delta_{nm}$.  
Because other contributions are not dominant in dc photocurrent generation, we only focus on $J^{({\rm I})}_{\rm intra} (t)$ and $J^{(\rm{II})}_{\rm intra} (t)$.  
Comparing with the total current $J(t)$ in Eq.~(\ref{eq:jtot_orb}), $J(\omega\!=\!0)$ in the ferroelectric EI reproduces the main features of $J^{(\rm{II})}_{\rm intra} (\omega\!=\!0)$ since the injection current contribution is dominant [see e.g., Fig.~5], and the sum of $J^{(\rm{I})}_{\rm intra} (\omega \!=\! 0)$ and $J^{(\rm{II})}_{\rm intra} (\omega\!=\!0)$ shows reasonable agreement with $J(\omega\!=\!0)$.

\subsection{Connection in the two-band model}
Here, we present the connection $\mathcal{A}_{nm}(k)$ in the two-orbital model. 
Using the eigenvector $\bm{U}_{\pm}(k)$ in Eqs.~(\ref{eq:eigenstates_1}) and (\ref{eq:eigenstates_2}), the intraband connection $\mathcal{A}_{\pm \pm}(k)$ is given by
\begin{align}
&\mathcal{A}_{\pm \pm}(k) = i\bm{U}^{\dag}_{\pm}(k)\frac{\partial }{\partial k} \bm{U}_{\pm}(k) .
\end{align}
Combining Eqs.~(\ref{eq:eigenstates_uk})-(\ref{eq:eigenstates_par}), we have 
\begin{align}
&\mathcal{A}_{\pm \pm}(k) =  - \frac{h_x(k) h'_y(k) - h_y(k) h'_x(k)}{2h(k) \left[ h(k) \pm h_z(k)\right]},
\label{eq:connectionintrapm}
\end{align}
where we defined $h(k) = |\bm{h}(k)|$ and $h'_{\mu}(k) = \partial_k h_{\mu}(k)$. 
In the same way, the interband connection is given by 
\begin{align}
\mathcal{A}_{+-}(k) = r_{+-}(k)= i\bm{U}^{\dag}_{+}(k)\frac{\partial }{\partial k} \bm{U}_{-}(k) .  
\end{align}
Using Eqs.~(\ref{eq:eigenstates_uk})-(\ref{eq:eigenstates_par}), we obtain
\begin{align}
\mathcal{A}_{+-}(k) 
&= \frac{h_x(k) h'_y(k) - h_y(k) h'_x(k)}{2 h(k) \sqrt{ h(k)^2-h_z(k)^2}} 
\notag \\
&-i \frac{ h(k) h'_z(k) - h_z(k) h'(k) }{2 h(k) \sqrt{ h(k)^2-h_z(k)^2}} .
\label{eq:connection_inter}
\end{align}
Using the pseudospin representation 
\begin{align}
&\left[
\begin{array}{cc}
\mathcal{A}_{++}(k) & \mathcal{A}_{+-}(k)
\\
\mathcal{A}_{-+}(k) & \mathcal{A}_{--}(k)
\end{array}
\right]
=\mathcal{A}_0(k) \sigma_0 + \bm{\mathcal{A}}(k)\cdot \bm{\sigma}, 
\label{eq:Berrysigma}
\end{align}
$\bm{\mathcal{A}}(k)=\left( \mathcal{A}_1(k), \mathcal{A}_2(k),\mathcal{A}_3(k) \right)$ is given by
\begin{align}
&\mathcal{A}_{1}(k)=  \frac{h_x(k) h'_y(k) - h_y(k) h'_x(k)}{2 h(k) \sqrt{ h(k)^2-h_z(k)^2}} ,
\\
&\mathcal{A}_{2}(k)= \frac{ h(k) h'_z(k) - h_z(k) h'(k) }{2 h(k) \sqrt{ h(k)^2-h_z(k)^2}},
\\
&\mathcal{A}_{3}(k)= \frac{h_z(k)}{2h(k)}\frac{h_x(k) h'_y(k) - h_y(k) h'_x(k) }{h(k)^2-h_z(k)^2}.
\end{align}

\subsection{Polarization}
In the ground state, the lower band is filled at all $k$, i.e., $\braket{\hat{d}^{\dag}_{k,-} \hat{d}_{k,-}}=1$ and $\braket{\hat{P}_{\rm inter}}=0$. 
Hence, combining Eqs.~(\ref{eq:Pintra}) and (\ref{eq:connectionintrapm}), the polarization $P=\braket{\hat{P}_{\rm intra}}/N$ is given by 
\begin{align}
P =  \int \! \frac{dk}{2\pi} \mathcal{A}_{--}(k) =  - \int \! \frac{dk}{2\pi}\frac{h_x(k) h'_y(k) - h_y(k) h'_x(k)}{2h(k) \left[ h(k) - h_z(k)\right]} ,
\end{align}
where the second term in Eq.~(\ref{eq:Pintra}) is zero due to $\braket{\hat{d}^{\dag}_{k,-} \hat{d}_{k',-}}=\delta_{k,k'}$~\citeSM{sipe2000SM}.  
Since the hybridization parameters are real (${\rm Im}[\phi_{\pm}]=0$) in the ground state, using Eqs.~(\ref{eq:hameq_x})-(\ref{eq:hameq_z}), we obtain 
\begin{align}
P = \int \! \frac{dk}{2\pi} \frac{V\phi_+(2t_{ab}+V\phi_-)}{4h(k) \left[ h(k) - h_z(k)\right]} . 
\end{align}
Because $h(k)= |\bm{h}(k)|$ and $h_z(k)=-2t_h\cos k + D + V\Delta n$ are even for $k$, $P\ne 0$ when $V\phi_+\ne 0$ and  $2t_{ab}+V\phi_- \ne 0$. 
Hence, the excitonic order parameter $\phi_{+}$ and interchain hopping $t_{ab}$ ($\phi_-$) are necessary for the spontaneous polarization. 

In the ground state of the model with $t_{ab} > 0$, the phase of the order parameter $\phi_+ = |\phi_+|e^{i\theta_+}$ is fixed at $\theta_+=0$ or $\pi$ because of the $Z_2$ symmetry breaking. 
Then, the polarization is given by $P>0$ at $\theta_+=0$ or $P < 0$ at $\theta_+=\pi$, indicating that the direction of the polarization is determined by the phase $\theta_+$. 
If the phase is switched from $\theta_+=0$ to $\pi$, associated with the change of the polarization direction, the direction of the second-order direct current is also inverted.

\subsection{Intraband current in the two-band model}
Employing the pseudospin representation $\tilde{\bm{\rho}}(k,t)=\left(\tilde{\rho}_1(k,t), \tilde{\rho}_2(k,t),\tilde{\rho}_3(k,t)\right)$ in the band basis,  the intraband current $J^{({\rm I})}_{\rm intra} (t)$ in Eq.~(\ref{eq:Jintra1_ap}) is given by 
\begin{widetext}
\begin{align}
J^{({\rm I})}_{\rm intra} (t) =\! -2 E(t) \! \int \! \frac{dk}{2\pi} \left[ \frac{\partial \mathcal{A}_1 (k,t) }{\partial k}  \tilde{\rho}_{1}(k,t) \!+\!  \frac{\partial \mathcal{A}_2 (k,t) }{\partial k} \tilde{\rho}_{2}(k,t)  \right] 
\! -4E(t) \! \int \! \frac{dk}{2\pi} \mathcal{A}_{3}(k,t) \left[ \mathcal{A}_{1}(k,t)\tilde{\rho}_{2}(k,t) \!-\! \mathcal{A}_{2}(k,t)\tilde{\rho}_{1}(k,t)   \right] . 
\end{align}
\end{widetext}
Since $\varepsilon_{\pm}(k,t) = \pm h(k,t)$, the current $J^{({\rm II})}_{\rm intra} (t)$ in Eq.~(\ref{eq:Jintra2_ap}) is simply given by 
\begin{align}
J^{({\rm II})}_{\rm intra} (t) =  2 \int \frac{dk}{2\pi}  \frac{\partial h(k,t)}{\partial k} \tilde{\rho}_{3}(k,t) . 
\label{eq:intracurrentII_b}
\end{align}

\subsection{Transformation of densities}
The densities $\tilde{\rho}_{mn}(k)=\braket{\hat{d}^{\dag}_{k,n} \hat{d}_{k,m}}$ in the band basis and $\rho_{\beta\alpha}(k)=\braket{\hat{c}^{\dag}_{k,\alpha} \hat{c}_{k,\beta}}$ in the orbital basis are related with $\tilde{\rho}_{mn}(k) = \sum_{\alpha,\beta} U_{\alpha n}(k) U^*_{\beta m}(k) \rho_{\beta \alpha}(k)$.  
Using the eigenvectors in Eqs.~(\ref{eq:eigenstates_1}) and (\ref{eq:eigenstates_2}), the densities in the band basis $\tilde{\bm{\rho}}(k,t)=\left( \tilde{\rho}_1(k,t), \tilde{\rho}_2(k,t),\tilde{\rho}_3(k,t) \right)$ are given by 
\begin{align}
\tilde{\rho}_1(k,t) =& -\frac{h_x(k,t)\rho_x(k,t) + h_y(k,t)\rho_y(k,t)}{\sqrt{h_x(k,t)^2+h_y(k,t)^2}} \frac{h_z(k,t)}{h(k,t)} 
\notag \\
&+ \sqrt{1-\frac{h_z(k,t)^2}{h(k,t)^2}} \rho_z(k,t), 
\\
\tilde{\rho}_2(k,t) =&-\frac{h_x(k,t)\rho_y(k,t) - h_y(k,t)\rho_x(k,t)}{\sqrt{h_x(k,t)^2+h_y(k,t)^2}}, 
\\
\tilde{\rho}_3(k,t) =& \frac{\bm{h}(k,t)\cdot\bm{\rho}(k,t)}{h(k,t)}, 
\label{eq:trans_rho3}
\end{align} 
where $\bm{\rho}(k,t)=\left( \rho_x(k,t), \rho_y(k,t),\rho_z(k,t)\right)$ is the density in the orbital basis.

\subsection{Nonlinear dc conductivity}
When the electric field with the frequency $\omega_p$ is applied,  the second-order photocurrent at $\omega=0$ is characterized by
\begin{align}
J(\omega=0) &=   \sigma_{xxx}(\omega=0;\omega_p,-\omega_p) E(\omega_p) E(-\omega_p) 
\notag \\
&+ \sigma_{xxx}(\omega=0;-\omega_p,\omega_p) E(-\omega_p) E(\omega_p).
\end{align}
In this paper, we employ the electric field $E(t) =E_0 \sin \omega_p t= E(\omega_p)e^{-i\omega_pt}+E(-\omega_p)e^{i\omega_pt}$. 
Hence, $E(\omega_p)^* = E(-\omega_p) = E_0/(2i)$ and 
\begin{align}
J(\omega=0) 
= \frac{1}{2}\sigma_{xxx}(\omega=0;\omega_p) E_0^2. 
\end{align} 
Here, we defined $2\sigma_{xxx}(\omega \!=\! 0;\omega_p) \! \equiv \! \sigma_{xxx}(\omega \!=\! 0;\omega_p,\!-\omega_p)+\sigma_{xxx}(\omega \!=\! 0;-\omega_p,\omega_p)$. 

In the real-time simulations, the intraband current  $({\rm I})$ after relaxation ($t\gg \gamma^{-1}$) behaves $J^{({\rm I})}_{\rm intra} (t)  \sim \sigma^{({\rm I})}_{xxx}(\omega=0;\omega_p) E(t)^2$. 
Hence, the nonlinear conductivity for the shift current  is defined  as
\begin{align}
\sigma^{({\rm I})}_{xxx}(\omega=0;\omega_p) = \frac{2}{E_0^2T_p} \int^{t_m+T_p}_{t_m} J^{({\rm I})}_{\rm intra} (t)  dt , 
\end{align} 
where $t_m \gg  \gamma^{-1}$  and $T_p = 2\pi /\omega_p$. 
In our actual calculation, we use $t_m = 10 / \gamma$ and $T_p \rightarrow N_p T_p$ with $N_p>10$ to take the average.    
In the same way, we estimate $\sigma^{({\rm II})}_{xxx}(0;\omega_p)$ for the injection current $J^{({\rm II})}_{\rm intra} (t) $. 

\begin{figure}[b]
\vspace{-1.5mm} 
\begin{center}
\includegraphics[width=\columnwidth]{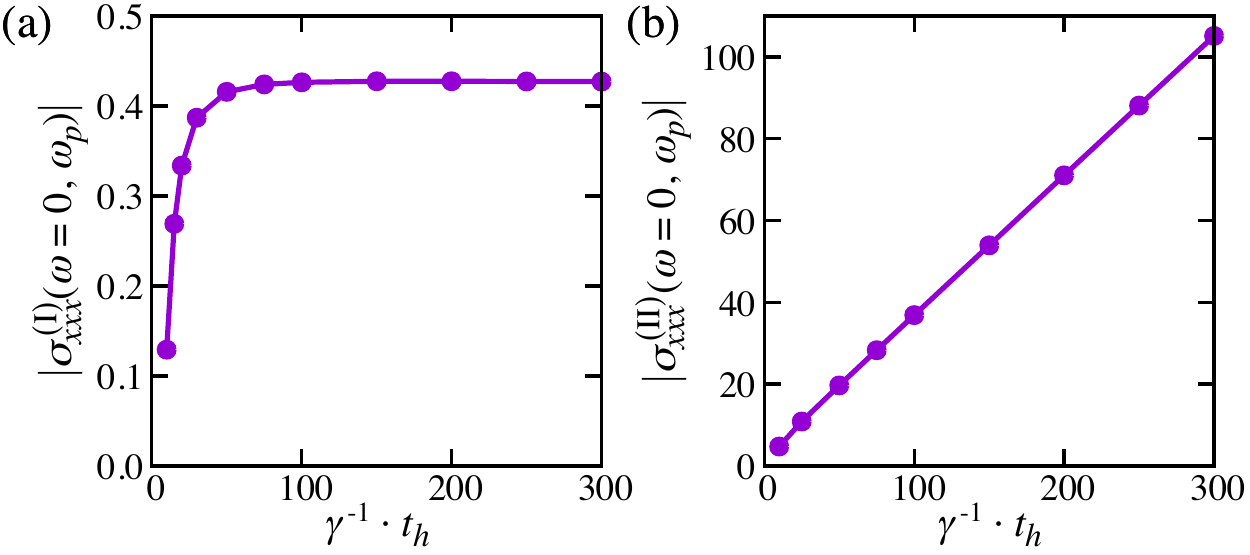}
\caption{
$\gamma$-dependence of the nonlinear conductivities of the (a) shift current $\sigma_{xxx}^{\rm (I)}(\omega=0;\omega_p)$ and (b) injection current $\sigma_{xxx}^{\rm (II)}(\omega=0;\omega_p)$ at $\omega_p/t_h=0.8$.  
$D/t_h=1$, $t_{ab}/t_h=0.2$, $V/t_h=1.1$, and $E_0/t_h = 0.0001$ are used. }
\label{sfig1}
\end{center}
\vspace{-5mm} 
\end{figure}

Here, we supplementarily show the $\gamma$ (relaxation term) dependence of the nonlinear dc conductivity  in the ferroelectric EI (FEI) state. 
Figure~\ref{sfig1}(a) and \ref{sfig1}(b) show $\sigma^{({\rm I})}_{xxx}(\omega=0;\omega_p)$ (shift current) and  $\sigma^{({\rm II})}_{xxx}(\omega=0;\omega_p)$ (injection current) at $\omega_p > E_g$ (band gap), respectively. 
In contrast to the shift current $\sigma^{({\rm I})}_{xxx}$, which converges at finite $\gamma^{-1}$, the injection current $\sigma^{({\rm II})}_{xxx}$ is linearly proportional to $\gamma^{-1}$.
This is because the time profile of the injection current generation is characterized by $dJ^{\rm (II)}_{\rm intra}(t)/dt$ and the value saturates to $J^{\rm (II)}_{\rm intra}(t)\propto \gamma^{-1}$ [see e.g., Fig.~4(a) in the main text] due to dissipative processes~\citeSM{sipe2000SM,zhang2019SM,holder2020SM}, which is qualitatively different from the shift current generation.
In this paper, the relaxation rate $\gamma$ in the EOM is introduced phenomenologically. 
It has the physical meaning of the scattering rate of the  photoexcited carriers, and may have disorder, phonon, and many-body contributions. 
The calculation of these rates from microscopic theory is an important open question.
Recently, in the noninteracting system, the correction of the formula of the injection current  has been proposed in the clean limit ($\gamma \rightarrow0$) by considering a fermionic bath in the Floquet Green's function approach~\citeSM{matsyshyn2021SM}. 
The proposed correction term is negligible when $\gamma \gg E_0$~\citeSM{matsyshyn2021SM} and does not change our main results because we assume $E_0 / \gamma \sim 0.01$ [except for the data at $\gamma=0$ in Fig.~4(a)].


\section{Velocity gauge}
In our main calculations, we employed the length gauge, in which the external field is introduced as 
\begin{align}
\hat{\mathcal{H}}_E(t) = &-\sum_{i,j}\sum_{\alpha,\beta} t_{i\alpha,j\beta} \hat{c}^{\dag}_{i,\alpha} \hat{c}_{j,\beta} + \hat{\mathcal{H}}_V 
\notag\\
& -E(t)\sum_{j,\alpha} R_{j,\alpha}  \hat{c}^{\dag}_{j,\alpha} \hat{c}_{j,\alpha}. 
\end{align}  
On the other hand, in the velocity gauge, the external field $A(t)$ (vector potential) is introduced via the Peierls substitution.
The covariant Hamiltonian~\citeSM{bukov2015SM} is  
\begin{align}
\hat{\mathcal{H}}_A(t) & =
 -\sum_{i,j}\sum_{\alpha,\beta} t_{i\alpha,j\beta} e^{iA(t) \left( R_{i,\alpha}-R_{j,\beta} \right)}  \hat{c}^{\dag}_{i,\alpha} \hat{c}_{j,\beta} + \hat{\mathcal{H}}_V 
\notag \\
& = \sum_{k} \sum_{\alpha,\beta} h^{(0)}_{\alpha\beta}\left( k-A(t) \right)  \hat{c}^{\dag}_{k,\alpha}  \hat{c}_{k,\beta} + \hat{\mathcal{H}}_V,  
\end{align} 
where $-\sum_{i,j} t_{i\alpha,j\beta}  \hat{c}^{\dag}_{i,\alpha} \hat{c}_{j,\beta}=\sum_{k}  h^{(0)}_{\alpha\beta}(k)  \hat{c}^{\dag}_{k,\alpha}  \hat{c}_{k,\beta}$.  
Because the interaction term is composed of $V\hat{n}_{i,\alpha}\hat{n}_{j,\beta}$, $\hat{\mathcal{H}}_V$ is invariant under the transformation. 
In this velocity gauge, the current operator is given by 
\begin{align}
\hat{J}_A(t)  = - \frac{\delta\hat{\mathcal{H}}_A(t) }{\delta A} = 
\sum_{k} \sum_{\alpha,\beta} v^{(0)}_{\alpha\beta}\left(k-A(t)\right) \hat{c}^{\dag}_{k,\alpha}  \hat{c}_{k,\beta}, 
\end{align}
with $v^{(0)}_{\alpha\beta}(k) = \partial_k h^{(0)}_{\alpha\beta}(k) $~\citeSM{scalapino1992SM,ventura2017SM}. 

Employing the HF approximation
\begin{align}
\hat{\mathcal{H}}_V \rightarrow \hat{\mathcal{H}}^{\rm HF}_V  = \sum_{k} \sum_{\alpha,\beta} h^{\rm(HF)}_{\alpha\beta}(k) \hat{c}^{\dag}_{k,\alpha}  \hat{c}_{k,\beta}, 
\end{align}
we have the tdMF Hamiltonian under the external field, 
\begin{align}
\hat{\mathcal{H}}^{\rm HF}_A(t)  = \sum_{k} \sum_{\alpha,\beta} \left[ h^{(0)}_{\alpha\beta}\left(k-A(t)\right) + h^{\rm (HF)}_{\alpha \beta}(k,t) \right] \hat{c}^{\dag}_{k,\alpha}  \hat{c}_{k,\beta}. 
\end{align}
In the pseudospin representation, the EOM in the tdMF for the particle density $\bm{\rho}_A(k,t)$ is given by  
\begin{align}
&\frac{\partial}{\partial t} \bm{\rho}_A(k,t) 
= 2\left[ \bm{h}^{(0)}\left( k \! - \! A(t) \right) + \bm{h}^{\rm(HF)}(k,t) \right]\times \bm{\rho}_A(k,t)  ,
 \label{eq:EOMVG}
\end{align} 
where 
\begin{align}
h^{(0)}_x(k) &= 0 , 
\notag \\
h^{(0)}_y(k) &=2t_{ab}  \sin  \frac{ k}{2}   , 
\notag \\
h^{(0)}_z(k) &= -2t_h\cos  k  + D , 
\label{eq:ham0VG}
\end{align} 
and 
\begin{align}
h^{\rm(HF)}_x(k,t) &= -V {\rm Re}[ \phi_+(t) ] \cos \frac{ k}{2}-  V{\rm Im}[\phi_-(t)] \sin \frac{ k}{2} , 
\notag \\
h^{\rm(HF)}_y(k,t) &=  -V {\rm Im}[ \phi_+(t) ] \cos \frac{ k}{2}+V {\rm Re}[\phi_-(t)] \sin  \frac{ k}{2}  ,
\notag \\
h^{\rm(HF)}_z(k,t) &=  V\Delta n(t). 
\label{eq:hamHFVG}
\end{align} 
The current (per unit) is given by 
\begin{align}
J_A(t)  = 2\int \frac{dk}{2\pi} \bm{v}^{(0)}\left(k-A(t)\right) \cdot \bm{\rho}_A(k,t)  
\label{eq:JVG}
\end{align}
with $\bm{v}^{(0)}(k) = \partial_k \bm{h}^{(0)}(k)$. 
If we assume $\hat{c}_{j,a}^{\dag} \hat{c}_{j,b} \hat{c}_{j,b}^{\dag} \hat{c}_{j,a}   =( \hat{c}_{j,a}^{\dag} \hat{c}_{j,b} e^{-iA(t)/2} ) ( \hat{c}_{j,b}^{\dag} \hat{c}_{j,a}  e^{iA(t)/2} )$ and define $\phi(x/2) \equiv \braket{ \hat{c}_{j,a}^{\dag} \hat{c}_{j,b} } e^{-i A(t)/2}$ in the Fock term, $k$ in the HF part is replaced as $\bm{h}^{\rm(HF)}(k,t) \! \rightarrow \! \bm{h}^{\rm(HF)}(k-A(t),t)$  [$\bm{v}^{(0)}(k)\rightarrow \partial_k \bm{h}(k,t) $],  but the phase factor $e^{\pm iA(t)/2}$ introduced in the order parameter does not change our final results. 

\begin{figure}[t]
\begin{center}
\includegraphics[width=\columnwidth]{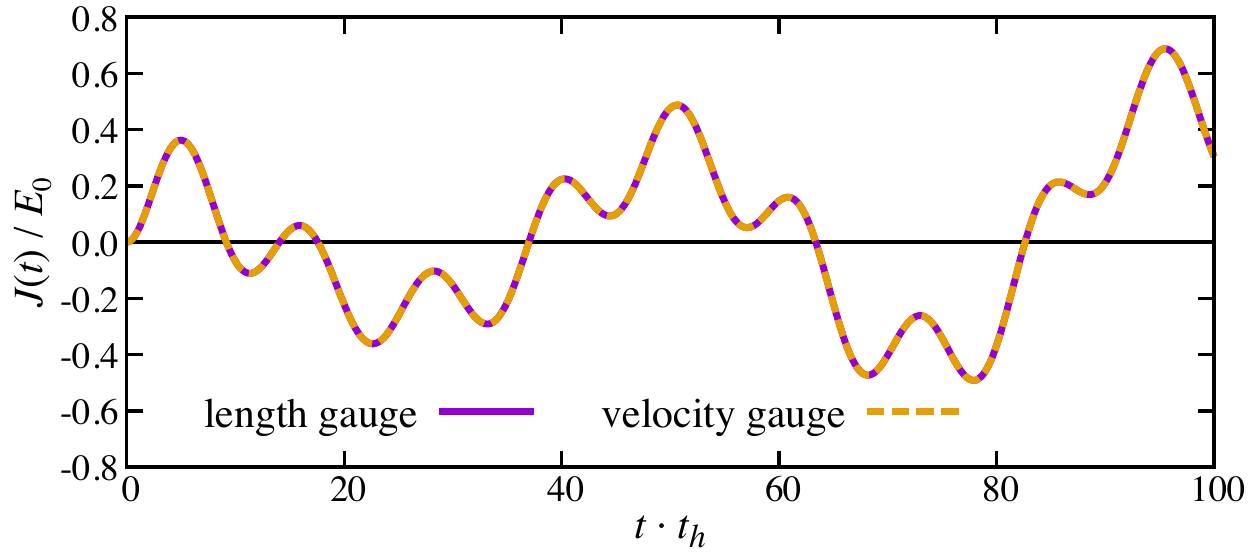}
\caption{
Time evolution of the total current $J(t)$ at $\gamma=0$ in the length gauge (solid line) and velocity gauge (dashed line), where $D/t_h=1$, $t_{ab}/t_h=0.2$, $V/t_h=1.1$, $E_0/t_h = 0.0001$, and $\omega_p/t_h=0.144$ are used. 
}
\label{sfig2}
\end{center}
\end{figure}

The calculated $J_A(t)$ in the FEI state is shown in Fig.~\ref{sfig2}, where we assume no relaxation ($\gamma=0$). 
Since $E(t) = -\partial_t A(t)$, we used $A(t) = E_0/\omega_p(\cos \omega_pt - 1)$ corresponding to $E(t) = E_0\sin \omega_pt$. 
In Fig.~\ref{sfig2}, we also plot the current $J(t)$ calculated in the length gauge, which is consistent with $J_A(t)$ in the velocity gauge. 
Therefore, we can obtain the consistent results between the length and velocity gauges. 

As pointed out in Ref.~\citeSM{passos2018SM}, when the relaxation term $\gamma[\bm{\rho}(k,t)-\bm{\rho}_{\rm eq}(k)]$ is introduced in the EOM, the perfect translation into the velocity gauge is not simple, where the transformation of the equilibrium distribution $\bm{\rho}_{\rm eq}(k)$ makes difficulty.  
Note that we can obtain qualitatively consistent results in our model even though the translation of the relaxation term is not perfect. 
In the analytical (diagrammatic) formulation of the response function, a different phenomenology has also been employed~\citeSM{passos2018SM,parker2019SM,kaplan2020SM,holder2020SM}, where each frequency is replaced as $\omega \rightarrow \omega + i\gamma$, which may also give the qualitatively consistent results.  

In the view of the real-time simulation, the amplitude of the vector potential $A(t)\propto E_0 /\omega_p$ (velocity gauge) diverges at $\omega_p \rightarrow 0$, which leads to numerical difficulties at the low-frequency regime.  
In terms of numerical estimation of conductivity using real-time simulations, the length gauge may be better than the velocity gauge, and we thus employed the length gauge in our numerical calculations.


\section{Origin of current injection}

\subsection{Action and injection current}
Here, employing the velocity gauge for the systematic perturbative expansion, we discuss the origin of the injection current. 

In the velocity gauge, the action for the EI state in the zigzag chain model may be given by 
\begin{widetext}
\begin{align}
S = \int \! \frac{dk}{2\pi}  \int \! d\tau \, \Bigl\{ \bar{\psi}(k,\tau) \left[ \partial_{\tau} \sigma_0 
+ \mathcal{H}_{0} (k-A(\tau)) +  \mathcal{H}^{\rm HF}_V (k,\tau) \right] \psi(k,\tau)
+ \frac{1}{2V}  |\Delta_+(\tau)|^2 + \frac{1}{2V} |\Delta_-(\tau)|^2 + \frac{V}{2} \Delta n(\tau)^2 \Bigr\},
\label{eq:effectiveaction}
\end{align}
\end{widetext}
where $\psi(k,\tau) = [c_{k,a}(\tau), c_{k,b}(\tau)]^{T}$ and $\tau$ is the imaginary time.  
$\mathcal{H}_0(k) = \bm{h}^{(0)}(k)\cdot \bm{\sigma}$ and $\mathcal{H}^{\rm HF}_V(k,t) = \bm{h}^{\rm (HF)}(k,t)\cdot \bm{\sigma}$, where $\bm{h}^{(0)}(k)$ and $ \bm{h}^{\rm (HF)}(k,t)$ are given in Eqs.~(\ref{eq:ham0VG}) and (\ref{eq:hamHFVG}), respectively. 
Here, we express the order parameter as $\Delta_{\pm}(t) = V\phi_{\pm}(t)$. 

For the perturbative analysis, we divide the Hamiltonian into the equilibrium and perturbative terms as  $\mathcal{H}_{0} (k-A(t)) +  \mathcal{H}^{\rm HF}_V (k,t)  = \mathcal{H}_{\rm eq} (k) +  \delta \Delta (k,t)$. 
In equilibrium, the hybridization parameters are real, i.e., ${\rm Im}[\Delta^{\rm eq}_{\pm}]=0$, and the Hamiltonian $\mathcal{H}_{\rm eq}(k) = \bm{h}(k)\cdot \bm{\sigma}$ is composed of 
\begin{align}
h_x(k) & =-\Delta^{\rm eq}_+ \cos \frac{ k}{2} , 
\notag \\
h_y(k) & =  (2 t_{ab} + \Delta^{\rm eq}_- )\sin\frac{ k}{2} ,
\notag \\
h_z(k) &= -2t_h \cos k + D + V \Delta n^{\rm eq}. 
\label{eq:Ham_eqgs}
\end{align} 
$\delta \Delta (k,t)$ is the difference from equilibrium given by $\delta \Delta (k,t) =  \delta \Delta_A (k,t) +  \delta \Delta_{\rm MF} (k,t)$, where the first term is $\delta \Delta_A (k,t) = \mathcal{H}_{0} (k-A(t))- \mathcal{H}_{0} (k)$ and the second term is due to the time-dependent  MF parameters, $\delta \Delta_{\rm MF} (k,t) =  \mathcal{H}^{\rm HF}_V(k,t) - \mathcal{H}^{\rm HF}_{V, {\rm eq}}(k)$.

Using the Fourier transformation, we have the action 
\begin{align}
S= \int \!  \frac{dk}{2\pi} \sum_{m,n} &\bar{\psi}(k,\omega_m) \bigl[ (-i\omega_n\sigma_0 + \mathcal{H}_{\rm eq}(k))\delta_{m,n} 
\notag \\
+&\delta\Delta(k,\Omega_{m-n}) \bigr] \psi(k,\omega_n) 
+ \left( \bm{\Delta}^2 \; {\rm terms} \right), 
\end{align}
where $\omega_n$ ($\Omega_n$) is the fermionic (bosonic) Matsubara frequency and $\bm{\Delta}$ indicates the MF parameters $\Delta_{\pm}$ and $\Delta n$. 
The diagonal term ($m=n$) corresponds to the bare fermion propagator, 
\begin{align}
G^{-1}_0(k,\omega_n) = i\omega_n\sigma_0 - \mathcal{H}_{\rm eq}(k). 
\end{align}
Integrating the fermion field~\citeSM{altland_simons_2010SM,sun2020junSM}, we obtain the action 
\begin{align}
S[A,\bm{\Delta}] =  -{\rm Tr} \ln  \left[\hat{G}_0^{-1} - \delta\hat{\Delta}\right] 
+ \left( \bm{\Delta}^2 \; {\rm terms} \right) , 
\end{align}
where ${\rm Tr}[\cdots]$ includes the $k$ and $\omega_n$ integrals. 
For  perturbative analyses, we expand the action with respect to $\delta \hat{\Delta}$, 
\begin{align}
S[A,\bm{\Delta}]  =  {\rm Tr} \ln  \hat{G}_0
\!+\! \sum_{M=1} \frac{1}{M}{\rm Tr} \left[ \left( \hat{G}_0 \delta\hat{\Delta} \right)^M\right] 
\!+\! \left( \bm{\Delta}^2 \; {\rm terms} \right) . 
\end{align}

The injection current is described by the loop triangle diagram [see Fig.~\ref{sfig4}]~\citeSM{parker2019SM}. 
Since the derivative of the action with respect to $A$ gives the current, the action $S^{(3)}_{\rm IC} =  {\rm Tr} [  \hat{G}_0 \delta\hat{\Delta}^{(1)}  \hat{G}_0 \delta\hat{\Delta}^{(1)}  \hat{G}_0 \delta\hat{\Delta}^{(1)}]/3$ gives the injection current, where $\delta\hat{\Delta}^{(1)} \propto A$ is the perturbation at first order. 
The external field term $\delta \Delta^{(1)}_A (k,t) = - \partial_k \mathcal{H}_0(k)  A(t)= -\mathcal{V}_0(k) A(t)$ is characterized by 
\begin{align}
\mathcal{V}_0(k)=\bm{v}^{(0)}(k) \cdot \bm{\sigma}=t_{ab}\cos (k/2) \sigma_2 + 2t_h\sin(k) \sigma_3.
\end{align} 
When the MF parameters are optically active in the linear response regime, we have to incorporate $\delta \Delta^{(1)}_{\rm MF}(k,\Omega)$. 
Writing the sum for $k$ and $\omega_n$ explicitly~\citeSM{altland_simons_2010SM}, the action $S^{(3)}_{\rm IC}$ and the injection current $J^{(2)}_{\rm IC}(0)$ may be given by 
\begin{widetext}
\begin{align}
&S^{(3)}_{\rm IC}
=  \frac{1}{3\beta} \! \sum_{l,m,n}  \int \!  \frac{dk}{2\pi} \, {\rm tr} \! \left[  G_0(k,\omega_n \!+\! \Omega_{l+m}) \delta \Delta^{(1)}(k,\Omega_l)  G_0(k,\omega_n \!+\! \Omega_m) \delta\Delta^{(1)}(k,\Omega_m)  
G_0(k,\omega_n) \delta \Delta^{(1)}(k,\!-\Omega_{l+m}) \right]  , 
\\
&J^{(2)}_{\rm IC}(0)  = \! - \frac{\delta S^{(3)}_{\rm IC}}{\delta A(\Omega_0)} 
= \!  \frac{1}{\beta} \! \sum_{m,n}  \int \!  \frac{dk}{2\pi}  {\rm tr} \! \left[ \mathcal{V}_0(k) G_0(k,\omega_n \!+\! \Omega_0) \delta \Delta^{(1)}(k,-\Omega_m)  G_0(k,\omega_n \!+\! \Omega_m) \delta\Delta^{(1)}(k,\Omega_m)  
G_0(k,\omega_n) \right]  ,
\label{eq:injection_der_SM}
\end{align}
\end{widetext}
respectively. 
Since we are considering the gapped system, the MF parameters are not optically active at $\Omega=0$ [i.e., $\delta \Delta^{(1)}_{\rm MF}(k,\Omega=0) = 0$], and we used $\frac{\delta}{\delta A(\Omega_0)}\delta\Delta^{(1)}(k,\Omega_m) = - \mathcal{V}_0(k)\delta_{m,0}$ in Eq.~(\ref{eq:injection_der_SM}). 

In our two-orbital model, the bare propagator $G_0(k,\omega_n)$ is given by
\begin{align}
&G_0(k,\omega_n) 
= \sum_{s=\pm} \frac{W_{s}(k)}{i\omega_n - \varepsilon_{s}(k)} , 
\label{eq:baregreensfunction} 
\end{align}
with  
\begin{align}
&W_{\pm}(k) = \frac{1}{2}\left[ \sigma_0 \pm \frac{\bm{h}(k)}{|\bm{h}(k)|}\cdot \bm{\sigma} \right]. 
\label{eq:Wpm}
\end{align}
Using this propagator, we can integrate $\omega_n$ and obtain 
\begin{widetext}
\begin{align}
J^{(2)}_{\rm IC}(0)  
=   -\sum_{m}\sum_{s_1,s_2, s_3 = \pm}   \int \!  \frac{dk}{2\pi} \, {\rm tr} \left[  \mathcal{V}_0(k)  W_{s_1} (k) \delta \Delta^{(1)}(k,-\Omega_m)  W_{s_2} (k)  \delta\Delta^{(1)}(k,\Omega_m)  W_{s_3} (k) \right]  &
\notag \\
\times 
\frac{1}{i\Omega_{0} - \varepsilon_{s_1}(k)+ \varepsilon_{s_3}(k)}
\left[ 
 \frac{f (\varepsilon_{s_2}(k))-f (\varepsilon_{s_3}(k))}{i\Omega_{m} - \varepsilon_{s_2}(k)+ \varepsilon_{s_3}(k)}
 + \frac{f (\varepsilon_{s_2}(k))-f (\varepsilon_{s_1}(k))}{i\Omega_{-m} - \varepsilon_{s_1}(k)+\varepsilon_{s_2}(k)}
 \right]&.
\end{align} 
\end{widetext}
The integrand shows divergence at $\varepsilon_{s_1}(k)= \varepsilon_{s_3}(k)$ due to $i\Omega_0 \rightarrow 0$. 
Because $\varepsilon_{\bar{s}}(k)=-\varepsilon_{s}(k)$ ($\bar{s} = -s$) in our two-orbital model, the leading contribution of the injection current is given by
\begin{widetext}
\begin{align}
J^{(2)}_{\rm IC}(0)  
\! \sim \! - \frac{1}{i\Omega_{0}} \! \sum_{m}\sum_{s = \pm}   \int \!  \frac{dk}{2\pi} \, {\rm tr} \left[  \mathcal{V}_0(k)  W_{s} (k) \delta \Delta^{(1)}(k,-\Omega_m)  W_{\bar{s}} (k)  \delta\Delta^{(1)}(k,\Omega_m)  W_{s} (k) \right]  
\left[ 
 \frac{1 \!-\! 2f (\varepsilon_{s}(k))}{i\Omega_{m} \!+\! 2\varepsilon_{s}(k)}
 \!+\! \frac{1 \!-\! 2f (\varepsilon_{s}(k))}{i\Omega_{-m} \!-\! 2\varepsilon_{s}(k)}
 \right] . 
\label{eq:injectioncurrent}
\end{align} 
\end{widetext}
When the relaxation term $\gamma$ is introduced, $i\Omega_0 \rightarrow 0 + i\gamma$ and $J^{(2)}_{\rm IC}(0)$ is proportional to $\gamma^{-1}$. 
Hence the injection current remains finite due to $\gamma$. 

For injection current, the trace
\begin{align}
{\rm tr} \left[  \mathcal{V}_0(k)  W_{s} (k) \delta \Delta^{(1)}(k,-\Omega)  W_{\bar{s}} (k)  \delta\Delta^{(1)}(k,\Omega)  W_{s} (k) \right]  
\label{eq:traceDelDel}
\end{align} 
is important in the $k$ integral in Eq.~(\ref{eq:injectioncurrent}).  
If this trace is odd for $k$, the integrands at $k$ and $-k$ cancel each other and thus $J^{(2)}_{\rm IC}(0)  =0$. 
For the nonvanishing injection current, this trace must not be an odd-$k$ function. 

If the MF parameters were not optically deformable at first order of $A(\Omega)$, the injection current vanishes. 
When $\delta \Delta^{(1)}_{\rm MF}(k,\Omega) = 0$, the trace (\ref{eq:traceDelDel}) is given by ${\rm tr} [  \mathcal{V}_0(k)  W_{s} (k) \mathcal{V}_0(k)   W_{\bar{s}} (k)  \mathcal{V}_0(k)  W_{s} (k) ] A(\Omega) A(-\Omega)$. 
Because 
\begin{align}
&W_s(-k) = W_s(k)^* = W_s(k)^T \\
&\mathcal{V}_0(-k) = -\mathcal{V}_0(k)^*= -\mathcal{V}_0(k)^T
\end{align} 
under time-reversal, we have
\begin{align}
&{\rm tr} \left[  \mathcal{V}_0(-k)  W_{s} (-k) \mathcal{V}_0(-k)  W_{\bar{s}} (-k) \mathcal{V}_0(-k)  W_{s} (-k) \right]  
\notag\\
&=- {\rm tr} \left[  \mathcal{V}_0(k)^T  W_{s} (k)^T \mathcal{V}_0(k)^T  W_{\bar{s}} (k)^T\mathcal{V}_0(k)^T  W_{s} (k)^T \right] 
\notag\\
&=- {\rm tr} \left[  \mathcal{V}_0(k)  W_{s} (k) \mathcal{V}_0(k)  W_{\bar{s}} (k) \mathcal{V}_0(k)  W_{s} (k) \right] , 
\label{eq:VWVWVWsym}
\end{align} 
where the symbol $T$ denotes the transposed matrix. 
Because of this relation, the integrands in Eq.~(\ref{eq:injectioncurrent}) at $k$ and $-k$ cancel each other. 
Therefore, the injection current within the independent particle approximation (IPA) vanishes due to the time-reversal symmetry. 

For nonvanishing injection current, $\delta \Delta^{(1)}_{\rm MF}(k,\Omega) \ne 0$ is necessary. 
When $\delta\Delta_{\pm}(t) =\Delta_{\pm}(t)-\Delta^{\rm eq}_{\pm}$ and $\delta n(t)=\Delta n(t)-\Delta n^{\rm eq}$ are nonzero, the total perturbation $\delta \Delta^{(1)}(k,\Omega)=\bm{F}(k,\Omega)\cdot \bm{\sigma}$ is given by
\begin{align}
F_x(k,\Omega) 
&= -\delta \Delta^{\rm R}_+(\Omega) \cos \frac{k}{2}-\delta \Delta^{\rm I}_-(\Omega) \sin \frac{k}{2}  ,
\notag \\
F_y(k,\Omega) 
&=- \! \left[  t_{ab} A(\Omega) \!+\! \delta \Delta^{\rm I}_+(\Omega) \right]\! \cos \frac{k}{2} + \delta \Delta^{\rm R}_-(\Omega) \sin \frac{k}{2},
\notag \\
F_z(k,\Omega) 
&=- 2t_h A(\Omega) \sin k +V \delta n(\Omega) , 
\label{eq:pert_SM}
\end{align} 
where the superscripts ${\rm R}$ and ${\rm I}$ indicate the real and imaginary part of the order parameter, respectively. 
When the MF parameters are nonzero, the trace (\ref{eq:traceDelDel}) can make even-$k$ contributions.  
For example, the following Eq.~(\ref{eq:Jinjection_eg}) has an even-$k$ contribution due to the excitonic order parameter $\delta \Delta^{\rm R}_{+}(\Omega)$. 
In this case, the integrands in Eq.~(\ref{eq:injectioncurrent}) at $k$ and $-k$ do not cancel each other. 
Thus, the injection current can be nonvanishing when the MF parameters are proportional to $A(\Omega)$.  
\begin{widetext}
\begin{align}
& {\rm tr} \left[  
\mathcal{V}_0(k) W_{s} (k) \Bigl( \left[t_{ab} A(-\Omega) + \delta\Delta^{\rm I}_+(-\Omega) \right]\cos \frac{k}{2} \sigma_2\Bigr)W_{\bar{s}} (k)  \Bigl(\delta\Delta^{\rm R}_+(\Omega)\cos \frac{k}{2} \sigma_1 \Bigr) W_{s} (k) \right]  
\notag \\
& = \frac{1}{4} s   \left( 2 t_{ab} + \Delta^{\rm eq}_- \right)^2 \Delta^{\rm eq}_+ \, t_{ab} \left( \sin \frac{k}{2} \right)^2 \left( \cos \frac{k}{2} \right)^4 \delta\Delta^{\rm R}_+(\Omega)\left[t_{ab} A(-\Omega) + \delta\Delta^{\rm I}_+(-\Omega) \right] + \cdots.
\label{eq:Jinjection_eg}
\end{align}
\end{widetext}

In noninteracting systems, time-reversal-symmetry breaking is required for a nonzero injection current under a linearly polarized light~\citeSM{zhang2019SM,ahn2020SM}. 
However, in our model, the ground-state Hamiltonian (\ref{eq:Ham_eqgs}) possesses the time-reversal symmetry $\mathcal{H}_{\rm eq}(-k)=\mathcal{H}_{\rm eq}(k)^*$. 
Hence,  in correlated electron systems, the broken time-reversal symmetry is not required at the equilibrium level.  
Instead, as indicated in Eq.~(\ref{eq:pert_SM}), the dynamical order parameter $\delta \Delta^{(1)}_{\rm MF}(k,\Omega)$ activated in the linear response regime plays the same role with time-reversal-symmetry breaking.  
Therefore,  in correlated electron systems, we can induce an injection current using an optically driven collective motion out of equilibrium. 

\begin{figure*}[t]
\begin{center}
\includegraphics[width=1.6\columnwidth]{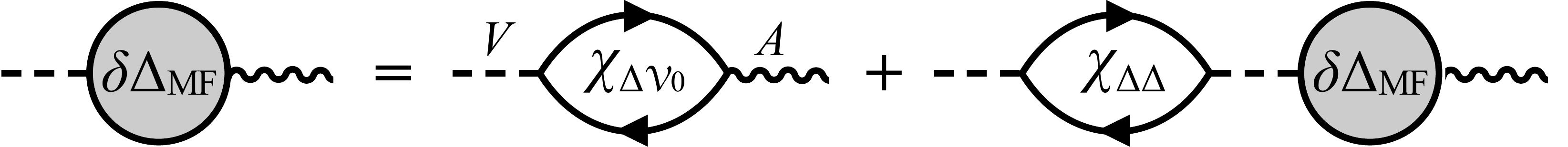}
\caption{
The diagrammatic representation of Eq.~(\ref{eq:selfeq}). 
The solid, wavy and dashed lines indicate the bare propagator, external field, and interaction, respectively.
}
\label{sfig3}
\end{center}
\end{figure*}

\subsection{Order parameter in the first-order perturbation}
For the nonvanishing injection current, we need $\delta \Delta_{\rm MF}(k,\Omega) \propto A(\Omega)$. 
In the main text, we have numerically shown the order parameter in the FEI  is optically active [see Fig.~2(b) and 2(c)]. 
In this section, we analytically show the relation $\delta \Delta_{\rm MF}(k,\Omega) \propto A(\Omega)$. 

The equation for the order parameter is given by $\Delta^{\rm R}_{+} = - (V/N) {\rm Tr} \bigl[\hat{L}_{\Delta^{\rm R}_+}\hat{G} \bigr]$, where $\hat{G}^{-1} = \hat{G}^{-1}_0 -\delta\hat{\Delta}$ and $L_{\Delta^{\rm R}_+} (k) = - \cos\frac{k}{2} \sigma_1$. 
Because we want to get $\delta \Delta_{+}(t) \propto A(t)$, here we expand the propagator as $\hat{G} = \hat{G}_0 + \hat{G}_0\delta\hat{\Delta}^{(1)} \hat{G}_0 + \cdots$.  
Considering the vector potential $A(t) = A(\Omega) e^{-i\Omega t} + {\rm c.c.}$ [$A(\Omega) = E(\Omega)/(i\Omega)$], we assume $\delta \Delta^{\rm R}_{+}(t) = \delta \Delta^{\rm R}_{+}(\Omega) e^{-i\Omega t} + {\rm c.c.}$.  
Using the first-order perturbative expansion, the order parameter away from equilibrium  is given by 
\begin{widetext}
\begin{align}
\delta \Delta^{\rm R}_{+}(\Omega_m) 
=  - \frac{V}{\beta} \sum_n \int \frac{dk}{2\pi} \, {\rm tr}  \Bigl[L_{\Delta^{\rm R}_+} (k) G_0(k,\omega_n+\Omega_m)
\delta \Delta^{(1)}(k,\Omega_m)G_0(k,\omega_n) \Bigr]. 
\label{eq:selfconsistent1st}
\end{align}
\end{widetext}
In the same way, we can obtain the equations for $\delta \Delta^{\rm I}_{+}(\Omega_m) $, $\delta\Delta^{\rm R}_{-}(\Omega_m)$, $\delta \Delta^{\rm I}_{-}(\Omega_m)$, and $\delta n(\Omega_m)$. 
Since $\delta \Delta^{(1)} (k,\Omega_m)$ includes the MF parameters, Eq.~(\ref{eq:selfconsistent1st}) corresponds to the self-consistent equation,  which composes the simultaneous equations with $\delta \Delta^{\rm I}_{+}(\Omega_m) $, $\delta \Delta^{\rm R}_{-}(\Omega_m)$, $\delta \Delta^{\rm I}_{-}(\Omega_m)$, and $\delta n(\Omega_m)$. 
Introducing the bare susceptibility, 
\begin{widetext}
\begin{align}
\chi^{(0)}_{M N}(\Omega_m)
= -  \frac{1}{\beta} \sum_n \int \frac{dk}{2\pi} \, {\rm tr}  \Bigl[ &L_M(k) G_0(k,\omega_n+\Omega_m) 
L_N(k) G_0(k,\omega_n) \Bigr] 
\label{eq:chiMN}
\end{align}
\end{widetext}
with
\begin{align}
L_{\mathcal{V}_0}(k) = &- \mathcal{V}_0(k) = -\bm{v}^{(0)}(k) \cdot \bm{\sigma}
\end{align}
and
\begin{align}
L_{M} (k) = &- \left[ \delta_{M,\Delta^{\rm R}_+} \cos\frac{k}{2}+  \delta_{M,\Delta^{\rm I}_-}   \sin\frac{k}{2} \right]\sigma_1 
\notag \\
&- \left[ \delta_{M,\Delta^{\rm I}_+} \cos\frac{k}{2} - \delta_{M,\Delta^{\rm R}_-} \sin\frac{k}{2} \right]\sigma_2 + \delta_{M,\delta n}\sigma_3, 
\end{align}
Eq.~(\ref{eq:selfconsistent1st}) becomes 
\begin{align}
\delta \Delta^{\rm R}_{+}(\Omega) 
=  V \chi^{(0)}_{\Delta^{\rm R}_{+} \mathcal{V}_0}(\Omega) A(\Omega)
+V \sum_{M} \chi^{(0)}_{\Delta^{\rm R}_{+}M}(\Omega)\delta M(\Omega) ,
\end{align}
where $M = \Delta^{\rm R}_+,  \Delta^{\rm I}_+ ,\Delta^{\rm R}_-, \Delta^{\rm I}_-, V\Delta n $. 
In the same way, we can derive the equations for $\delta \Delta^{\rm I}_{+}(\Omega) $, $\delta \Delta^{\rm R}_{-}(\Omega)$, $\delta \Delta^{\rm I}_{-}(\Omega)$, and $\delta n(\Omega)$. 
To describe the simultaneous equations compactly, we introduce the vector 
\begin{align}
\delta \bm{\Delta}^{(1)}_{\rm MF} (\Omega)  \! =\!  \! 
\left[
\delta \Delta^{\rm R}_{+} \! (\Omega) 
\;
\delta \Delta^{\rm I}_{+} \! (\Omega) 
\;
\delta \Delta^{\rm R}_{-} \! (\Omega) 
\;
\delta \Delta^{\rm I}_{-} \! (\Omega) 
\;
 V\delta n (\Omega) 
\right]^T \! ,
\end{align}
and 
\begin{align}
&\chi^{(0)}_{M \mathcal{V}_0}(\Omega) = \left[ \bm{\chi}^{(0)}_{\Delta \mathcal{V}_0} (\Omega) \right]_{M},
\\
&\chi^{(0)}_{MN}(\Omega) = \left[ \chi^{(0)}_{\Delta \Delta}(\Omega)  \right]_{MN},
\end{align}
for $M, N = \Delta^{\rm R}_+,  \Delta^{\rm I}_+ ,\Delta^{\rm R}_-, \Delta^{\rm I}_-, V\Delta n$, 
where $\bm{\chi}^{(0)}_{\Delta \mathcal{V}_0} (\Omega)$ is a vector and $ \chi^{(0)}_{\Delta \Delta}(\Omega)$ is a 5$\times$5 matrix. 
Using the above vectors and matrix, the simultaneous equations are summarized as 
\begin{align}
&\delta \bm{\Delta}^{(1)}_{\rm MF} (\Omega) 
=   V \bm{\chi}^{(0)}_{\Delta \mathcal{V}_0}(\Omega) A(\Omega) +  V \chi^{(0)}_{\rm \Delta \Delta}(\Omega)  \delta \bm{\Delta}^{(1)}_{\rm MF}  (\Omega). 
\label{eq:selfeq} 
\end{align}
This is the self-consistent equation for all MF parameters. 
Diagrammatically, Eq.~(\ref{eq:selfeq}) is expressed as Fig.~\ref{sfig3}~\citeSM{tsuji2015SM}. 
This self-consistent equation gives 
\begin{align}
\delta \bm{\Delta}^{(1)}_{\rm MF}(\Omega)
=   \frac{V \bm{\chi}^{(0)}_{\Delta \mathcal{V}_0}(\Omega)}{ I -  V \chi^{(0)}_{\rm \Delta \Delta}(\Omega)}  A(\Omega). 
\label{eq:oaop_1st}
\end{align}
Since the denominator includes $\chi^{(0)}_{\rm \Delta \Delta}(\Omega)$, $\delta \bm{\Delta}^{(1)}_{\rm MF}(\Omega)$ reflects the structure of the dynamical correlation function of the excitonic order parameter. 

\begin{figure*}[t]
\begin{center}
\includegraphics[width=1.69\columnwidth]{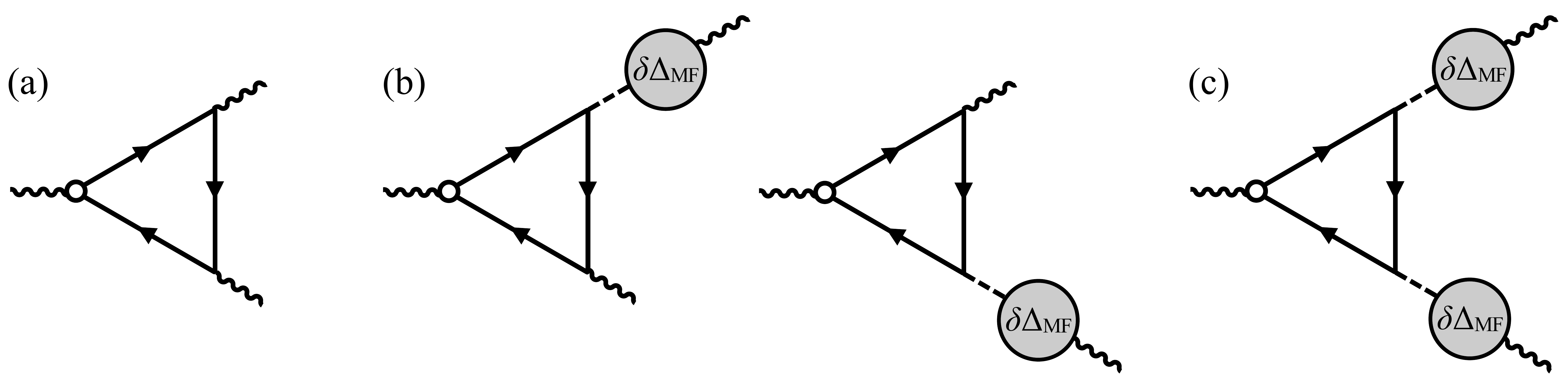}
\caption{
The diagrammatic representation of the injection current with the optically active MF parameters. 
(a) corresponds to the diagram in the IPA. 
(b) and (c) are diagrams with one collective mode and two collective modes, respectively. 
}
\label{sfig4}
\end{center}
\end{figure*}

To be $\delta \bm{\Delta}^{(1)}_{\rm MF} (\Omega) \ne 0$, the bare susceptibility $\bm{\chi}^{(0)}_{\Delta \mathcal{V}_0}(\Omega)$ must be nonzero. 
Using Eq.~(\ref{eq:baregreensfunction}), we obtain 
\begin{widetext}
\begin{align}
\chi^{(0)}_{M \mathcal{V}_0}(\Omega_m)
= \! \! \sum_{s_1,s_2 = \pm} \int \! \frac{dk}{2\pi} \,  {\rm tr}  \left[L_M(k) W_{s_1}\! (k) L_{\mathcal{V}_0}(k)  W_{s_2} \! (k) \right] 
 \frac{f (\varepsilon_{s_1}(k))- f(\varepsilon_{s_2}(k))}{i\Omega_m - \varepsilon_{s_1}(k)+ \varepsilon_{s_2}(k)}. 
\label{eq:chiMV0}
\end{align}
\end{widetext}
Here, the important contribution in the $k$ integral is ${\rm tr} \left[L_M(k) W_{s}(k) L_{\mathcal{V}_0}(k)  W_{\bar{s}}  (k) \right]$, which is proportional to
\begin{widetext}
\begin{align}
{\rm tr} \left[ \sigma_{\mu} W_{s} (k)L_{\mathcal{V}_0}(k) W_{\bar{s}} (k)  \right]
=
\left[ \tilde{\bm{h}}(k)\cdot\bm{v}^{(0)}(k)\right]\tilde{h}_{\mu}(k)
-v^{(0)}_{\mu}(k) - i s \left[\tilde{\bm{h}}(k)\times\bm{v}^{(0)}(k)\right]_{\mu} , 
\label{eq:sigWLVW}
\end{align}
\end{widetext}
where we defined $\tilde{\bm{h}}(k)=\bm{h}(k)/|\bm{h}(k)|$. 
In our two-orbital model [$v^{\rm (0)}_{x}(k)=0$], $ \tilde{\bm{h}}(k)\cdot\bm{v}^{(0)}(k) =\tilde{h}^{\rm (o)}_y(k)v^{\rm (e)}_y(k)+\tilde{h}^{\rm (e)}_z(k)v^{\rm (o)}_z(k)$ is odd for $k$. 
Here, the superscripts $\rm (e)$ and $\rm (o)$ indicate the even and odd functions for $k$, respectively. 
In the FEI ($\Delta_+^{\rm eq}\ne 0$ and $t_{ab}\ne 0$), we have 
\begin{widetext}
\begin{align}
&{\rm tr} \left[ \sigma_1 W_{s} (k)L_{\mathcal{V}_0}(k)   W_{\bar{s}} (k) \right]
 = 
\left[ \tilde{\bm{h}}(k)\cdot\bm{v}^{(0)}(k)\right]^{\rm (o)}\tilde{h}^{\rm (e)}_{x}(k)
- i s \left[\tilde{h}^{\rm (o)}_{y}(k)v^{\rm (o)}_{z}(k)- \tilde{h}^{\rm (e)}_{z}(k)v^{\rm (e)}_{y}(k) \right] 
,
\label{eq:sig1WLVW} \\
&{\rm tr} \left[ \sigma_2 W_{s} (k)L_{\mathcal{V}_0}(k)  W_{\bar{s}} (k)  \right]
=
\left[ \tilde{\bm{h}}(k)\cdot\bm{v}^{(0)}(k)\right]^{\rm (o)}\tilde{h}^{\rm (o)}_{y}(k) 
-v^{\rm {(e)}}_y(k) + i s  \tilde{h}^{\rm (e)}_{x}(k)v^{\rm (o)}_{z}(k) 
, 
\label{eq:sig2WLVW} \\
&{\rm tr} \left[ \sigma_3 W_{s} (k) L_{\mathcal{V}_0}(k)   W_{\bar{s}} (k) \right]
=
\left[ \tilde{\bm{h}}(k)\cdot\bm{v}^{(0)}(k)\right]^{\rm (o)}\tilde{h}^{\rm (e)}_{z}(k)
-v^{\rm {(o)}}_z(k)- i s \tilde{h}^{\rm (e)}_{x}(k)v^{\rm (e)}_{y}(k) 
.
\label{eq:sig3WLVW} 
\end{align} 
\end{widetext}
Because they include both even-$k$ and odd-$k$ contributions, the integrands with  ${\rm tr}  \left[L_M(k) W_{s_1}\! (k) L_{\mathcal{V}_0}(k)  W_{s_2} \! (k) \right]$ in Eq.~(\ref{eq:chiMV0}) at $k$ and $-k$ do not cancel each other. 
Hence,  in the FEI state, $\bm{\chi}^{(0)}_{\Delta \mathcal{V}_0}(\Omega)$ can be nonzero, and thus the MF parameters in $\delta \bm{\Delta}^{(1)}_{\rm MF}(\Omega)$ is proportional to $A(\Omega)$. 
This is consistent with the results in our numerical time-dependent calculations. 

Notice that the Fourier coefficient of the order parameter, e.g., $\delta \Delta^{\rm R}_{+}(\Omega)$, is complex in general.  
Defining 
\begin{align}
\delta \Delta^{\rm R}_{+}(\Omega) 
= |\delta \Delta^{\rm R}_{+}(\Omega)|e^{-i\varphi^{\rm R}_{+}}, 
\end{align}
the real part of the time-dependent order parameter  $\delta \Delta^{\rm R}_{+}(t) = \delta \Delta^{\rm R}_{+}(\Omega) e^{-i\Omega t} + \delta \Delta^{\rm R}_{+}(\Omega)^* e^{i\Omega t}$ [$\Delta^{\rm R}_{+}(-\Omega)=\Delta^{\rm R}_{+}(\Omega)^*$] is given by  
\begin{align}
\delta \Delta^{\rm R}_{+}(t) 
 = 2|\delta \Delta^{\rm R}_{+}(\Omega)| \cos \left( \Omega t + \varphi^{\rm R}_{+} \right), 
\end{align}
where $\varphi^{\rm R}_{+}$ brings the phase shift on the time-dependent order parameter. 

Including the optically deformable MF parameters, the perturbation at first order is given by $\delta \Delta^{(1)}(k,\Omega) = \delta \Delta^{(1)}_{A}(k,\Omega) + \delta \Delta^{(1)}_{\rm MF}(k,\Omega)$, and the injection current in Eq.~(\ref{eq:injection_der_SM}) is composed of the contributions described diagrammatically  in Fig.~\ref{sfig4}. 
The loop triangle diagram in Fig.~\ref{sfig4}(a) corresponds to the injection current within the IPA, which vanishes in our model [due to Eq.~(\ref{eq:VWVWVWsym})].     
When one photon input is modified by the order parameter as shown in Fig.~\ref{sfig4}(b), this contribution can lead to the nonvanishing injection current [see e.g., Eq.~(\ref{eq:Jinjection_eg})].  
Because $\delta \bm{\Delta}^{(1)}_{\rm MF}(\Omega)$ in Eq.~(\ref{eq:oaop_1st}) reflects the structure of the corrected susceptibility $\bm{\chi}^{}_{\Delta \mathcal{V}_0}(\Omega)= [ I -  V \chi^{(0)}_{\rm \Delta \Delta}(\Omega)]^{-1}\bm{\chi}^{(0)}_{\Delta \mathcal{V}_0}(\Omega)$, the conductivity $\sigma_{xxx}$ of the injection current exhibits two peaks  at the sub-band-gap collective mode frequencies.     
The injection current in Fig.~\ref{sfig4}(c) driven by two modified photon inputs can also be nonzero [see e.g., Eq.~(\ref{eq:Jinjection_eg})].  
When two inputs are the same order parameter, for example, the contribution proportional to $\delta \Delta^{\rm R}_+(\Omega)\delta \Delta^{\rm R}_+(-\Omega)$ vanishes due to the $k$-symmetry of the integrand in Eq.~(\ref{eq:injectioncurrent}). 
Hence, when two order parameters contribute, the injection current is proportional to two different order parameters, e.g., $\delta \Delta^{\rm R}_{+}(\Omega)\delta \Delta^{\rm I}_{+}(-\Omega) + {\rm c.c.} = 2|\delta \Delta^{\rm R}_{+}(\Omega)| |\delta \Delta^{\rm I}_{+}(\Omega)| \cos (\varphi^{\rm R}_+-\varphi^{\rm I}_+)$. 
However, due to the phase difference [e.g., $\varphi^{\rm R}_+-\varphi^{\rm I}_+$], the injection current driven by two order parameters may tend to be suppressed.

\subsection{Absence of injection current in the non-ferroelectric EI ($t_{ab}=0$)}
When $t_{ab}=0$, the EI state is not ferroelectric ($P=0$). 
In this case, Eqs.~(\ref{eq:sig1WLVW})-(\ref{eq:sig3WLVW}) are given by 
\begin{align}
&{\rm tr} \left[ \sigma_1 W_{s} (k)L_{\mathcal{V}_0}(k)   W_{\bar{s}} (k) \right]
 = \tilde{h}^{\rm (e)}_{x}(k) \tilde{h}^{\rm (e)}_z(k)v^{\rm (o)}_z(k) ,
\notag\\
&{\rm tr} \left[ \sigma_2 W_{s} (k)L_{\mathcal{V}_0}(k)  W_{\bar{s}} (k)  \right]
= i s  \tilde{h}^{\rm (e)}_{x}(k)v^{\rm (o)}_{z}(k) , 
\notag \\
&{\rm tr} \left[ \sigma_3 W_{s} (k) L_{\mathcal{V}_0}(k)   W_{\bar{s}} (k) \right]
={\tilde{h}^{\rm (e)}_z(k)}^2v^{\rm (o)}_z(k) -v^{\rm {(o)}}_z(k) ,
\notag
\end{align} 
and all of them are odd for $k$. 
Therefore, integrands with ${\rm tr}  [L^{\rm (e)}_M(k) W_{s_1}\! (k) L_{\mathcal{V}_0}(k)  W_{s_2} \! (k) ]$ vanishes, and $\delta \Delta_+(\Omega)$ and $\delta n(\Omega)$ that are even for $k$ cannot be activated by $A(\Omega)$. 
On the other hand, $\delta \Delta_-(\Omega)$ that are odd for $k$ can be activated by $A(\Omega)$. 
In this case,  because 
\begin{align}
\delta \Delta^{(1)}(k,\Omega) =& -\delta \Delta^{\rm I}_-(\Omega) \sin \frac{k}{2}\sigma_1
+\delta \Delta^{\rm R}_-(\Omega) \sin \frac{k}{2}\sigma_2
\notag \\
&- 2t_h A(\Omega) \sin k \sigma_3, 
\end{align} 
we find
\begin{align}
\delta \Delta^{(1)}(-k,\Omega)
= - \delta \Delta^{(1)}(k,\Omega). 
\notag 
\end{align} 
When $t_{ab}=0$, because $W_{s} (-k)=W_{s} (k)$ and $\mathcal{V}_0(-k) =-\mathcal{V}_0(k) $, we have 
\begin{align}
&{\rm tr} \! \left[ \! \mathcal{V}_0(\!-k)  W_{s} (-k) \delta \Delta^{(1)}\! (\!-k,\! \Omega)  W_{\bar{s}} (\!-k)  \delta\Delta^{(1)}\!(\!-k,\! \Omega')  W_{s} (\!-k) \! \right] 
\notag \\
&=-{\rm tr} \! \left[  \mathcal{V}_0(k)  W_{s} (k) \delta \Delta^{(1)}(k,\Omega)  W_{\bar{s}} (k)  \delta\Delta^{(1)}(k,\Omega')  W_{s} (k) \right]. 
\label{eq:integrak-k}
\end{align} 
Therefore, the injection current $J^{(2)}_{\rm IC}(0) $ in Eq.~(\ref{eq:injectioncurrent}) vanishes in the non-ferroelectric EI state at $t_{ab}=0$.

\subsection{Vertex correction of shift current}
Here, we supplementally discuss the vertex correction of the shift current contribution. 
The shift current contribution may be derived from the action $S^{(3)}_{\rm ShC} =  {\rm Tr} [  \hat{G}_0 \delta\hat{\Delta}^{(1)}  \hat{G}_0 \delta\hat{\Delta}^{(2)}]$~\citeSM{parker2019SM}, where $\delta\hat{\Delta}^{(2)} \propto A^2$ is the perturbation at second order. 
The bare electric field term is $\delta \Delta^{(2)}_A (k,t) =  [\partial^2_k \mathcal{H}_0(k)]  A(t)^2/2= \mathcal{T}_0(k) A(t)^2/2$. 
Here, we assume $\delta \Delta^{(1)}_{\pm} (t) \gg \delta \Delta^{(2)}_{\pm} (t)$ and focus simply on the contribution derived from $\delta \Delta^{(2)}_{A}(k,t)$.  
When $\delta \Delta^{(2)}=\delta \Delta^{(2)}_A$, as in Eq.~(\ref{eq:injection_der_SM}), the derivative of the action with respect to $A$ may give a shift current contribution at $\Omega_m$ $(> 0)$, 
\begin{align}
J^{(2)}_{\rm ShC}(0;\Omega_m) \!  = \! - \frac{1}{\beta} \! \sum_{n}  \! \! \int  \!  \frac{dk}{2\pi}  & {\rm tr}  \bigl[ \mathcal{T}_0(k)A(-\Omega_m) G_0(k,\omega_n \!+\! \Omega_m) 
\notag \\
& \times \delta\Delta^{(1)}(k,\Omega_m)  G_0(k,\omega_n) \bigr] 
\notag \\
 &+ [\Omega_m \leftrightarrow -\Omega_m] .
\end{align}
Since $\delta \Delta^{(1)} (k,\Omega) =  \delta \Delta^{(1)}_A (k,\Omega) +  \delta \Delta^{(1)}_{\rm MF} (k,\Omega)$, combining Eqs.~(\ref{eq:chiMN}) and (\ref{eq:oaop_1st}), the current $J^{(2)}_{\rm ShC}(0;\Omega)  =  \chi^{}_{ \mathcal{T}_0 \mathcal{V}_0} (\Omega) A(\Omega) A(-\Omega) + [\Omega \leftrightarrow -\Omega] $ is characterized by the response function 
\begin{align}
\chi^{}_{ \mathcal{T}_0 \mathcal{V}_0} (\Omega) =\chi^{(0)}_{ \mathcal{T}_0 \mathcal{V}_0} (\Omega) 
+ \bm{\chi}^{(0)}_{ \mathcal{T}_0 \Delta}(\Omega) \! \cdot \! \frac{V\bm{\chi}^{(0)}_{ \Delta \mathcal{V}_0}(\Omega) }{I \! - \! V\chi^{(0)}_{ \Delta \Delta}(\Omega)} ,
\label{eq:shiftVC}
\end{align}
where $\chi^{(0)}_{ \mathcal{T}_0 \mathcal{V}_0} (\Omega)$ [$\chi^{}_{ \mathcal{T}_0 \mathcal{V}_0} (\Omega)$] is a scalar. 
The first term $\chi^{(0)}_{ \mathcal{T}_0 \mathcal{V}_0} (\Omega)$ corresponds to the response function in the IPA, which can be nonzero at $\Omega > E_g$ (above-band-gap). 
The second term in Eq.~(\ref{eq:shiftVC}) is the correction term caused by the excitonic interaction $V$, which is equivalent to the solution of the Bethe-Salpeter equation with the vertex correction.   

This correction term leads to two consequences in the response function. 
First, since the correction term includes the susceptibility $\chi^{(0)}_{ \Delta \Delta}(\Omega)$ in the denominator, the pole in $[I -  V\chi^{(0)}_{ \Delta \Delta}(\Omega)]^{-1}$ gives rise to the resonant enhancement of the response, where the pole positions correspond to the collective mode frequencies. 
In Fig.~3(c) in the main text, we actually find the resonant peaks in $\sigma^{\rm (I)}_{xxx}$ in the tdMF while it is absent in the result in the IPA. 
Second, the correction term modifies the shape of the above-band-gap response function from the IPA.   
In Fig.~3(c), the many-body correction leads to the sign change of $\sigma^{\rm (I)}_{xxx}$ at $\Omega > E_g$. 
The diagonalized $I -  V\chi^{(0)}_{ \Delta \Delta}(\Omega)$ at $\Omega < E_g$ monotonically decreases with increasing $\Omega$ and crosses zero from positive to negative at the collective mode frequency ($\Omega_c$). 
The negative contribution in the correction term in Eq.~(\ref{eq:shiftVC}) at $\Omega > E_g$ $(> \Omega_c$) reduces the spectral weight from $\chi^{(0)}_{ \mathcal{T}_0 \mathcal{V}_0} (\Omega)$ in the IPA.  
Hence, because of the corrections derived from the order parameter dynamics, the shift current contribution $\sigma^{\rm (I)}_{xxx}$ in the tdMF is modified from the result in the IPA.


\section{Electron-phonon system}
\begin{figure}[b]
\begin{center}
\includegraphics[width=0.70\columnwidth]{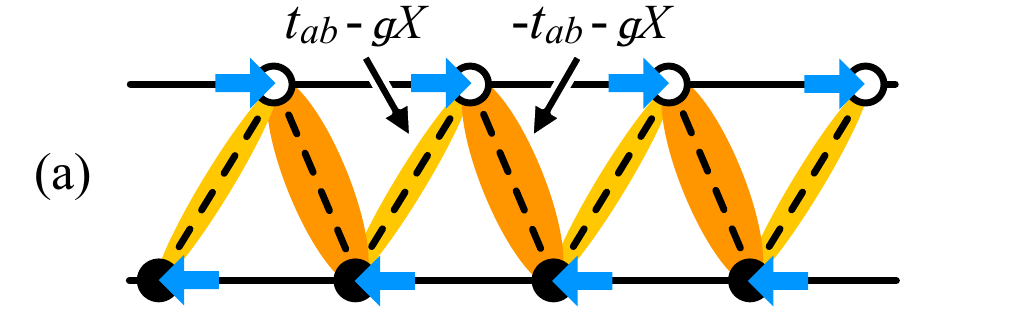}
\includegraphics[width=0.95\columnwidth]{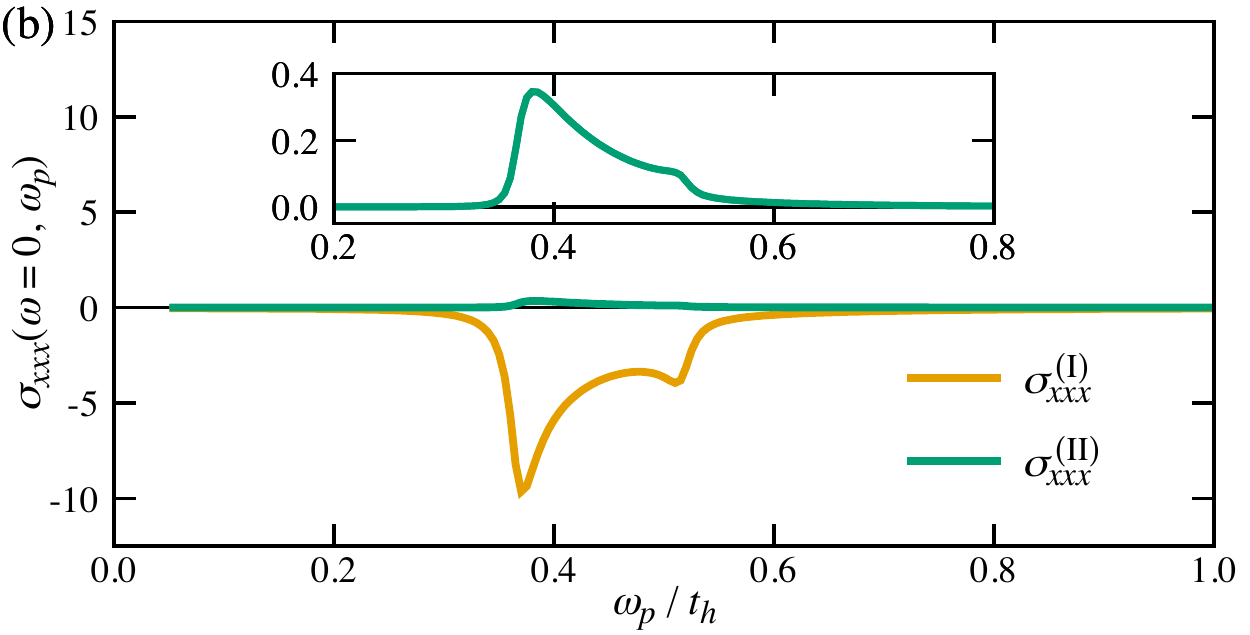}
\caption{
(a) Schematic picture of the lattice displacement in the electron-phonon coupled system. 
(b) Nonlinear conductivities $\sigma_{xxx}^{\rm (I)}(\omega=0;\omega_p)$ (shift) and $\sigma_{xxx}^{\rm (II)}(\omega=0;\omega_p)$ (injection) estimated by the time-dependent calculation, where $D/t_h=1.8$, $t_{ab}/t_h=0.2$, $\omega_0/t_h=0.01$, $\lambda/t_h=0.2$ ($\lambda=g^2/\omega_0$), $E_0/t_h = 0.0001$, and $\gamma/t_h=0.01$ are used. 
Inset is the enlarged view of $\sigma_{xxx}^{\rm (II)}(0;\omega_p)$.  
}
\label{sfig5}
\end{center}
\end{figure}

Here, we discuss the shift and injection currents in an electron-phonon coupled system that breaks the inversion symmetry [see Fig.~\ref{sfig5}(a)]. 

When the lattice displacement $u_{\alpha}$ is much smaller than the lattice constant (i.e., $u_{\alpha} \ll R_{j+1,\alpha} - R_{j,\alpha}$), the tight-binding Hamiltonian is approximately given by 
\begin{align}
\hat{\mathcal{H}}_{\rm el} = -\sum_{i,j}\sum_{\alpha,\beta} \left[ t_{i\alpha,j\beta} + t'_{i\alpha,j\beta} (u_{\alpha}- u _{\beta})\right]\hat{c}^{\dag}_{i,\alpha} \hat{c}_{j,\beta},  
\end{align}  
where $t'_{i\alpha, j\beta}$ is the first derivative of the transfer integral with respect to $R =R_{i,\alpha} - R_{j,\beta}$,  and  $t'_{i\alpha, j\beta} = -t'_{j\beta,i\alpha}$. 
Corresponding to the zigzag chain model in the main text, we define $-t_{ja,ja} = t_{jb,jb}= D$, $t_{j+1\alpha,j\alpha} = t_{\alpha}$ with $t_a=-t_b = t_h$, and $t_{ja,jb} = -t_{ja,j-1b}= t_{ab}$. 
Here, assuming the phonon mode shown in Fig.~\ref{sfig5}(a), we define $u_a = - u_b  = u/2 = X/(2\sqrt{M\omega_0})$ and $t'_{j a, jb} = -  t'_{j b, j a} = g\sqrt{M\omega_0}$, where $M$ and $\omega_0$ are the effective mass and frequency of the phonon mode at $q=0$. 
Because we are considering the interchain hopping  $t_{ja,jb} = -t_{ja,j-1b}= t_{ab}$, its derivative is given by $t'_{j a, jb} =  t'_{j a, j-1b} =g\sqrt{M\omega_0}$. 
Hence, we consider
\begin{align}
\hat{\mathcal{H}}_{\rm el} =
&- \sum_{j,\alpha} t_{\alpha} \left( \hat{c}^{\dag}_{j+1,\alpha} \hat{c}_{j,\alpha} \!+\! {\rm H.c.} \right)
+ D \sum_{j}  \left( \hat{n}_{j,a} \!-\! \hat{n}_{j,b} \right)
 \notag \\
&- \left( t_{ab} +g X \right)  \sum_j \left( \hat{c}^{\dag}_{j,a} \hat{c}_{j,b} \!+\! {\rm H.c.} \right) 
\notag \\
&+  \left( t_{ab} -  g X \right)\sum_{j}  \left( \hat{c}^{\dag}_{j,a} \hat{c}_{j-1,b} \!+\! {\rm H.c.}  \right).
\end{align}
Since this Hamiltonian has the same form as the Rice-Mele model~\citeSM{rice1982SM,fregoso2017SM}, shift current can be generated. 
The phonon system is described by 
\begin{align}
\mathcal{H}_{\rm ph}= N \left( \frac{p^2}{2M} + \frac{1}{2}M\omega_0^2 u^2\right) = \frac{1}{2}N\omega_0 \left( P^2 +  X^2 \right), 
\end{align}
where $p = \sqrt{M\omega_0}P$ ($u = X/\sqrt{M\omega_0}$) is the momentum (displacement) of the phonon mode at $q=0$.   
Employing the length gauge, the external field $E(t)$ is introduced as 
\begin{align}
\hat{\mathcal{H}}_E(t) 
&=-E(t) \sum_{j,\alpha} \left( R_{j,\alpha} + u_{\alpha} \right) \hat{n}_{j,\alpha}. 
\label{eq:hamE}
\end{align}
Because we are considering a weak electric field and a small lattice displacement ($u_{\alpha} \ll 1$), here we neglect the contribution from $E(t)u_{\alpha}$ for simplicity.  
We have numerically confirmed that the contribution from $E(t)u_{\alpha}$ does not change the results qualitatively. 

Incorporating the phonon dynamics, i.e., $X \rightarrow X(t)$, we consider the time-dependent Hamiltonian $\hat{\mathcal{H}}(t) = \hat{\mathcal{H}}_{\rm el}(t) + \mathcal{H}_{\rm ph}(t) + \hat{\mathcal{H}}_E(t)$.  
Here, the EOM for the electron system is given by 
\begin{align}
\frac{\partial}{\partial t} \bm{\rho}(k,t) 
&= 2\bm{h}(k,t) \times \bm{\rho}(k,t)  
-  E(t) \frac{\partial}{\partial k} \bm{\rho}(k,t) 
\notag \\
&-   \gamma \left[\bm{\rho}(k,t) - \bm{\rho}_{\rm eq}(k)\right] 
\label{eq:EOM_elph_el}
\end{align}
with 
\begin{align}
h_x(k,t) &= -2g X(t)  \cos \frac{ k}{2}  , 
\notag \\
h_y(k,t) &=2t_{ab}  \sin  \frac{ k}{2},
\notag \\
h_z(k,t) &= -2t_h\cos k + D , 
\label{eq:HAM_elph_pspin}
\end{align} 
and the EOM for the phonon is given by
\begin{align}
\frac{\partial^2 X(t)}{\partial t^2} =  -\omega_0^2 X(t) + g \omega_0 \phi_{\rm el}(t)  
\label{eq:EOM_elph_ph}
\end{align}
with $\phi_{\rm el}(t) =  \frac{1}{N}\sum_j [  \braket{\hat{c}^{\dag}_{j,a} \hat{c}_{j,b}}(t) +  \braket{\hat{c}^{\dag}_{j,a} \hat{c}_{j-1,b} }(t) + {\rm c.c.} ]$. 

Here, we solve the EOM (\ref{eq:EOM_elph_el}) and (\ref{eq:EOM_elph_ph}) simultaneously and calculate the nonlinear conductivity $\sigma_{xxx}(\omega=0,\omega_p)$ from the time-dependent intraband currents $J^{({\rm I})}_{\rm intra} (t)$ and $J^{({\rm II})}_{\rm intra} (t)$ [see Eqs.~(\ref{eq:Jintra1_ap}) and (\ref{eq:Jintra2_ap})].  
Figure~\ref{sfig5}(b) shows $\sigma_{xxx}(\omega=0,\omega_p)$, where we assume the phonon frequency $\omega_0$ is much smaller than the band gap $E_g$, corresponding to realistic systems. 
Note that, because expensive (long-time) simulations are necessary for reliable accuracy in the low frequency region, we plot the data above $\omega_p/t_h = 0.05$, which is larger than the phonon frequency $\omega_0 /t_h = 0.01$ we use here. 
In Fig.~\ref{sfig5}(b), $\sigma^{\rm (I)}_{xxx}(\omega=0,\omega_p)$ exhibits the nonzero shift current contribution as we expected. 
Although the injection current contribution $\rm (II)$ is also nonzero, its value is much smaller than the shift current contribution. 

The nonzero injection current may be caused by the phonon motion introduced via $X(t)$ in Eq.~(\ref{eq:HAM_elph_pspin}).  
The EOM (\ref{eq:EOM_elph_ph}) implies 
\begin{align}
\delta X(\Omega)=  -\frac{g \omega_0}{\Omega^2 - \omega^2_0} \delta \phi_{\rm el}(\Omega), 
\end{align}
where $\delta X(\Omega)$ and $\delta \phi_{\rm el}(\Omega)$ are the Fourier coefficients of $\delta X(t) =X(t)- X^{\rm eq}$ and $\delta \phi_{\rm el}(t) = \phi_{\rm el}(t) - \phi^{\rm eq}_{\rm el}$, respectively. 
When $\delta X(\Omega) \propto \delta \phi_{\rm el}(\Omega) \propto E(\Omega)$ in the ferroelectric state ($X^{\rm eq} \ne 0$), the phonon motion $X(t)$ in Eq.~(\ref{eq:HAM_elph_pspin}) plays a similar role with the real part of the excitonic order parameter. 
However, in the realistic condition $E_g \gg \omega_0$, 
\begin{align}
\delta X(\Omega\sim E_g) \sim  -\frac{g \omega_0}{E^2_g} \delta \phi_{\rm el}(E_g) \ll 1
\end{align}
at $\Omega \sim E_g$, indicating that $\delta X(\Omega)$ is very small  in the above-band-gap regime.    
Hence, the impact of the phonon motion on the electronic system is limited in the above-band-gap regime due to the energy scale mismatch between the phonon mode and electronic band gap.
Therefore, the injection current contribution caused by $\delta X(\Omega)$ should be very weak. 
Because the contribution from the dynamical phonon is small, $\sigma^{\rm (I)}_{xxx}(\omega=0,\omega_p)$ (shift current) shows good agreement with the conductivity obtained by the independent particle approximation.


\section{Next-nearest-neighbor interchain hopping $t_{ab}'$}

\begin{figure}[b]
\begin{center}
\includegraphics[width=\columnwidth]{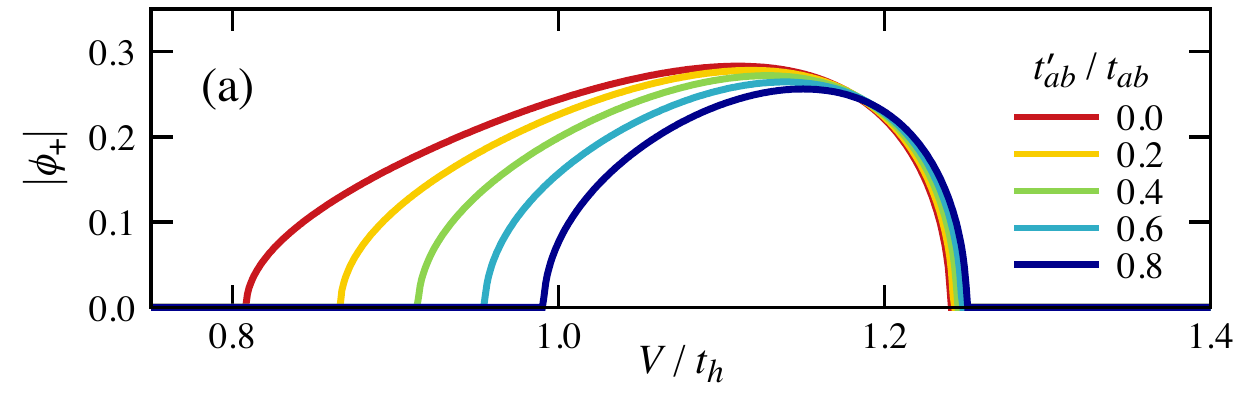}
\\
\includegraphics[width=\columnwidth]{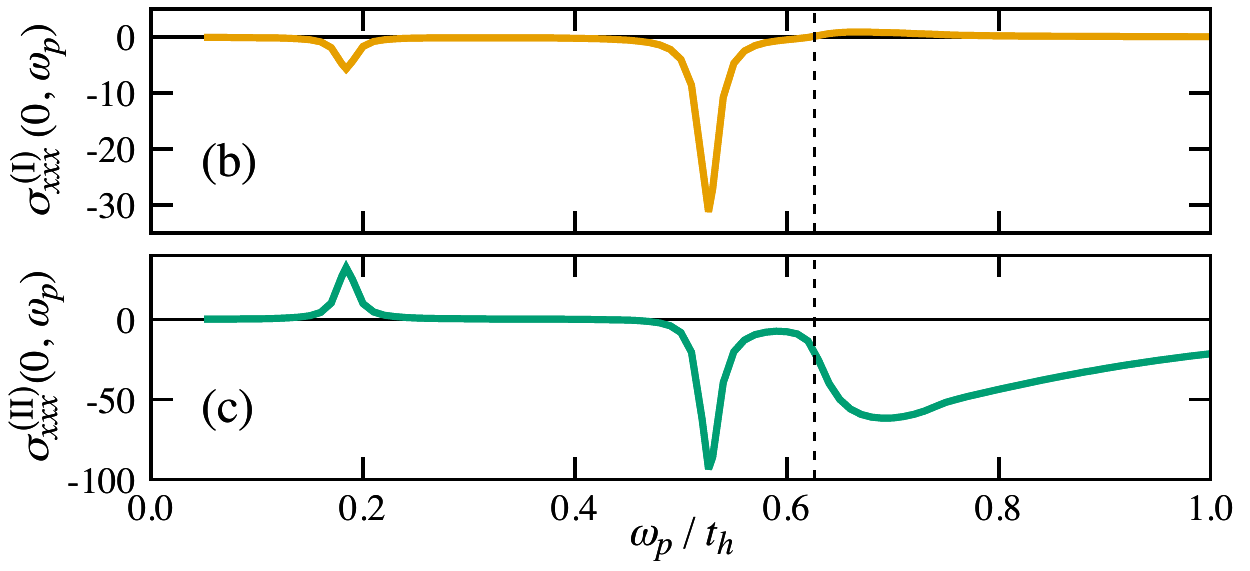}
\caption{(a) $V$ and $t_{ab}'$ dependence of the order parameter $\phi_+$ in the ground (FEI) state, where $D/t_h \!=\! 1$ and  $t_{ab}/t_h \!=\! 0.1$. 
(b)  $\sigma_{xxx}^{\rm (I)}(\omega=0;\omega_p)$ (shift) and (c) $\sigma_{xxx}^{\rm (II)}(\omega=0;\omega_p)$ (injection) at $t_{ab}'/t_{ab}=0.8$ and $V/t_h=1.1$. 
The black dashed line indicates the band gap. 
$E_0/t_h = 0.0001$ and $\gamma/t_h=0.01$ are used.
}
\label{sfig6}
\end{center}
\end{figure}

Here, we discuss the effect of the next-nearest-neighbor (NNN) interchain hopping 
\begin{align}
\hat{\mathcal{H}}_{ab}' =
&- t_{ab}'  \sum_{j}   \left( \hat{c}^{\dag}_{j+1,a} \hat{c}_{j,b} + {\rm H.c.} \right) 
\notag \\
&+ t_{ab}'  \sum_{j}  \left( \hat{c}^{\dag}_{j,a} \hat{c}_{j-2,b} + {\rm H.c.}  \right)
\end{align}
on the BPVE in the FEI. 
In the pseudospin representation, this NNN hopping is introduced by $h_y(k,t) \rightarrow h_y(k,t) + 2t_{ab}' \sin (3k/2)$ in Eq.~(3) in the main text. 

Figure~\ref{sfig6}(a) shows the order parameter $\phi_+$ (in equilibrium) with the NNN hopping $t_{ab}'$. 
Even when $t_{ab}'$ is nonzero, the EI state is stabilized and  the phase of the order parameter $\phi_+=|\phi_+|e^{i\theta_+}$ is fixed at $\theta_+ =0$ or $\pi$. 
Similar to the result in Fig.~1(b) in the main text, the region of the FEI phase is suppressed with increasing the interchain hopping $t_{ab}'$. 
Figures~\ref{sfig6}(b) and \ref{sfig6}(c) show the nonlinear response functions $\sigma_{xxx}^{\rm (I)}(\omega=0;\omega_p)$ (shift current) and $\sigma_{xxx}^{\rm (II)}(\omega=0;\omega_p)$ (injection current), respectively, in the FEI phase with $t_{ab}'\ne 0$. 
For comparison with the results at $t_{ab}=0.2 t_h$ ($t_{ab}'=0$) in the main text, we set $t_{ab} + t_{ab}' = 0.18 t_h$ with $t_{ab}'/t_{ab}=0.8$ ($t_{ab}=0.1t_h$), which gives roughly the same band gap energy. 
As shown in Fig.~\ref{sfig6}, even if $t_{ab}'$ is comparable to $t_{ab}$, $\sigma_{xxx}(\omega=0;\omega_p)$ retains the main features and the magnitudes of the shift and injection current is not strongly suppressed comparing with the results at $t_{ab}'=0$ in the main text. 
Therefore, the effect of the NNN hoping, which is anticipated in real materials, is minor on the BPVE in the FEI.

\end{appendix}


\bibliographystyleSM{apsrev4-1}
\bibliographySM{References_arXiv}

\end{document}